\documentclass[twocolumn,preprintnumbers,superscriptaddress]{revtex4-2}
\usepackage{amsmath}
\usepackage{amssymb}
\usepackage{graphicx}
\usepackage{hyperref}
\usepackage{physics}
\usepackage{xcolor}
\usepackage{xspace,slashed}
\usepackage{qcircuit}
\usepackage{multirow}
\usepackage[utf8]{inputenc}
\usepackage{subfigure}

\usepackage{array}

\newcommand{\PDF}{{\rm qPDF}_i(x, Q_0, \theta)}

\begin{document}

\title{
 Determining the proton content with a quantum computer}

\preprint{TIF-UNIMI-2020-30}

\newcommand{\MIaff}{TIF Lab, Dipartimento di Fisica, Universit\`a degli Studi di
  Milano and INFN Sezione di Milano, Milan, Italy.}

\newcommand{\TII}{Quantum Research Centre, Technology Innovation Institute, Abu Dhabi, UAE.}

\newcommand{\CERNaff}{CERN, Theoretical Physics Department, CH-1211
  Geneva 23, Switzerland.}

\newcommand{\UB}{Departament de F\'isica Qu\`antica i Astrof\'isica and Institut de Ci\`encies del Cosmos (ICCUB), Universitat de Barcelona, Barcelona, Spain.}

\newcommand{\BSC}{Barcelona Supercomputing Center, Barcelona, Spain.}

\author{Adri\'an P\'erez-Salinas}
\affiliation{\BSC}
\affiliation{\UB}
\author{Juan Cruz-Martinez}
\affiliation{\MIaff}
\author{Abdulla A. Alhajri}
\affiliation{\TII}
\author{Stefano Carrazza}
\affiliation{\MIaff}
\affiliation{\CERNaff}
\affiliation{\TII}

\begin{abstract}
    We present a first attempt to design a quantum circuit for the determination
    of the parton content of the proton through the estimation of parton
    distribution functions (PDFs), in the context of high energy physics (HEP).
    The growing interest in quantum computing and the recent developments of new
    algorithms and quantum hardware devices motivates the study of methodologies
    applied to HEP. In this work we identify architectures of variational
    quantum circuits suitable for PDFs representation (qPDFs). We show
    experiments about the deployment of qPDFs on real quantum devices, taking into
    consideration current experimental limitations. Finally, we perform a global
    qPDF determination from collider data using quantum computer simulation on
    classical hardware and we compare the obtained partons and related
    phenomenological predictions involving hadronic processes to modern PDFs.
\end{abstract}

\maketitle

\section{Introduction}

Quantum computing is a new computation paradigm that exploits the laws of
quantum mechanics to provide new strategies for addressing problems that are
nowadays considered to be difficult. The first quantum algorithms showing any
advantage over their classical counterparts date from the 1990s, being Shor's
algorithm for integer factorization and Grover's search the most
prominent ones~\cite{shor,grover}. During the last decade, we have witnessed an
impressively fast development of quantum computing, both for theoretical work
and hardware implementation perspectives. Nevertheless, currently existing
quantum devices are not powerful enough to run competitive quantum algorithms,
with respect to the state of the art of the classical ones.

Recent achievements such as {\sl quantum supremacy}~\cite{supremacy} have
introduced the so-called Noisy Intermediate-Scale Quantum (NISQ)
stage~\cite{nisq}. NISQ devices suffer from errors due to decoherence, noisy
gates and erratic read-out measurements, and thus, its performance is limited.
However, even at this early stage, quantum technologies may provide useful tools
for a broad range of applications. On the one hand, some standard fully
determined algorithms are well suited for NISQ
processors~\cite{cerveralierta:2018ising,suba:2019spectroscopy,bravyi:2018shallow,Bravyi:2020noisyshallow,ramoscalderer:2020unary}. 
In particular, there also exist some examples of quantum algorithms designed to address some problems in high energy physics (HEP)~\cite{hep_amplitudes-bepari2020, hep_simulation-bauer2019, hep_gluon-alexandru2019, hep_parton-lamm2020}.
On the other hand, the approach usually taken to harness the computational power
of these imperfect machines is based on hybrid methods combining quantum and
classical resources. For example, variational algorithms can be created whose
purpose is to optimize some quantity encoding a solution for a specific problem.
Among the great variety of quantum variational algorithms it is possible to find
examples in quantum
chemistry~\cite{peruzzo:2014vqe,higgott:2019vqe,aspuru:2005molecular,hempel:2018chemistry,jones:2019vqe},
quantum
simulation~\cite{li:2017simulation,kokail:2019simulation,cirstoiu:2020simulation},
combinatorial optimization~\cite{farhi:2014qoao}, solving linear systems of
equations~\cite{bravoprieto:2020linear,xu:2019linear,huang:2019linear} and state
diagonalization
~\cite{LaRose:2019diagonalization,BravoPrieto:2020diagonalization}. Some of
these examples are already characterized as Quantum Machine Learning (QML)
applications, based on variational~\cite{perezsalinas:2020reuploading,
mitarai:2018circuit,Zhu:2019circuit,Schuld:2020circuit,lloyd:2020embeddings} and
non-variational~\cite{liu:2020svm,Rebentrost:2014svm,lloyd:2013ml} approaches.
Furthermore, QML is a field that is expected to surpass the current performance
and ubiquity of classical Machine Learning (ML) when the current limitations of
quantum devices will be overcome.

The QML approach to quantum computing is an interesting research topic which can
be adapted and tested on research problems already addressed by ML techniques.
Motivated by this idea, we propose to investigate the possibility to use
quantum computing for the determination of parton distribution functions (PDFs).
In perturbative QCD, PDFs are used to describe the non-perturbative structure of
hadrons~\cite{Butterworth:2015oua,Forte:2020yip}. These functions are typically
determined by means of a supervised regression model which compares a wide set
of experimental data with theoretical predictions computed with a PDF
parametrization.

In this work we first propose the most suitable QML architecture for PDFs
representation and then perform experiments about its deployment on real quantum
devices, taking into account the current experimental limitations. Then, we
adapt the NNPDF
methodology~\cite{Carrazza:2019mzf,AbdulKhalek:2019ihb,AbdulKhalek:2019bux,Ball:2018iqk,Ball:2017nwa,Ball:2014uwa,Ball:2012cx},
based on ML techniques, to operate in a QML environment, replacing
Neural-Networks with quantum circuits.

The novel quantum circuit parametrization for PDFs, that we call qPDFs in
the next paragraphs, follows the quantum model described in
Ref.~\cite{perezsalinas:2020reuploading}. The model is constructed as a
Parameterized Quantum Circuit (PQC) whose inner parameters depend both on PDF
data and trainable parameters. A PQC whose parameters are trainable is known as
a Variational Quantum Circuit (VQC). The circuit is applied to an initial quantum
state, for instance the ground state $\ket 0$, and the output state contains
information on PDFs. The determination of the circuit parameters is done with
standard classical optimization methods, using a predefined cost function.

There are different reasons for attempting a qPDFs determination. First, quantum
computing is expected to have a reduced energy consumption when compared to an
equivalent classical computer, and thus, we may expect saving power and reducing
its environmental impact. Secondly, as we show in this work, the number of
parameters needed to obtain an acceptable PDF fit is in average lower with
quantum models in comparison to modern PDF models. Furthermore, the
qPDF approach may take advantage from quantum entanglement, since the potential
outstanding power of quantum computing emerges from there. Finally, quantum
hardware may bring performance improvements in terms of running time for this
model when compared to the standard ML approach since the number of operations
needed to obtain an acceptable solution is lower and the model has an exact
hardware representation. On the other hand, we consider the qPDF model presented
in this work as proof-of-concept for future implementations, given that the
performance of quantum simulation on classical hardware and the stability of
real quantum device measurements are not competitive with the ML tools used by
modern PDF determinations.

The paper is structured as follows. Sec.~\ref{sec:qcpdf} provides an overall
description of the quantum circuit model for PDFs, while in
Sec.~\ref{sec:implementation} we identify its best architecture. In
Sec.~\ref{sec:quantumhardware} we discuss about the deployment of qPDFs on real
quantum devices. In Sec.~\ref{sec:lhc} we integrate the qPDF model in the NNPDF
fitting framework and perform a first global qPDF determination using data from experiments such as Tevatron or LHC.
In Sec.~\ref{sec:pheno} we compute Higgs observable predictions using the qPDF
fit. Finally, in Sec.~\ref{sec:conclusion} we present our conclusion and future
development directions.

\section{Quantum circuits for PDFs}\label{sec:qcpdf}

Quantum circuits are mathematically defined as operations acting on an initial
quantum state. Quantum computing usually makes use of quantum states constructed
out of qubits, that is, binary states represented as $\ket{\psi} = \alpha \ket 0
+ \beta \ket 1$. The states of a quantum circuit are commonly defined by
its number of qubits $n$, and, in general, the initial state of the circuit
$\ket{\psi_0}$ is the zero state $\ket 0^{\otimes n}$. A quantum circuit
implements an inner unitary operation $\mathcal U$ to the initial state
$\ket{\psi_0}$ to transform it into the final output state $\ket{\psi_f}$. For
some algorithms, this $\mathcal U$ gate is fully determined~\cite{shor, grover},
while other algorithms define its inner operation by means of some fixed
structure, so-called {\sl Ansatz}, and tunable parameters $\mathcal
U(\theta)$~\cite{peruzzo:2014vqe, bravoprieto:2020linear, xu:2019linear}. Those
are known as Parameterized Quantum Circuits (PQC). This kind of circuits is
useful in the NISQ era of quantum computing, since they provide a great
flexibility and allow to approximate unitary operations up to arbitrary
precision~\cite{kitaev1997, dawson:2005solovay}. The parameters defining the
PQCs can be trained using an optimization procedure known as a Variational
Quantum Circuit (VQC). It is possible then to use classical computational
resources to find the optimal configuration of a quantum circuit.

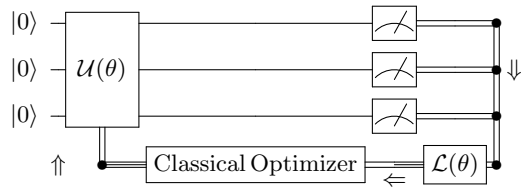
\begin{figure}[t]
\centering
\[
  \Qcircuit @R=0.5em @C=0.3em{
    \lstick{\ket 0} & \qw & \multigate{2}{\mathcal U(\theta)} & \qw & \meter & \cw & \control \cw \\
    \lstick{\ket 0} & \qw & \ghost{\mathcal U(\theta)} & \qw & \meter & \cw & \control \cwx \cw & \push{\Downarrow} \\
    \lstick{\ket 0} & \qw & \ghost{\mathcal U(\theta)} & \qw & \meter & \cw & \control \cwx \cw \\
     & \Uparrow & \control \cwx & \gate{\rm Classical\,Optimizer} \cw & \dstick{\Leftarrow} \cw & \gate{\mathcal{L}(\theta)} \cw & \control \cwx \cw \\
    }
\]
\caption{Operational scheme of a Variational Quantum Circuit. A unitary gate
$\mathcal U$, depending on some parameters $\theta$, transforms the initial
$\ket 0$ state into some output state. This state is measured and used to
compute a loss function $\mathcal{L}(\theta)$. The classical optimizer performs
an update on the parameters to minimize the value of $\mathcal{L}(\theta)$. New
parameters are then sent to the quantum circuit and the loop starts again.}
\label{fig:VQC}
\end{figure}

A VQC follows roughly three steps to solve a given problem, as schematically
shown in Fig.~\ref{fig:VQC}. First, a PQC $\mathcal U(\theta)$ is constructed
using a small set of single- and two-qubit parametric gates. The Ansatz of such
circuit may follow a particular path exploiting the special features of the
problem, or may also be a general one. After the Ansatz is applied to the
circuit, we must perform some measurements on the output quantum state to
extract information. Those measurements are used to evaluate a loss function
$\mathcal{L}(\theta)$ encoding the problem. The loss function should reach its
minimum as the problem is perfectly solved. The loss function $\mathcal
L(\theta)$ is passed to a classical optimizer that looks for the value
\begin{equation} \theta^*= {\rm argmin} \left(\mathcal L(\theta)\right).
\end{equation}
Classical optimizers need several function evaluations, thus when modifying the
set of parameters $\theta$ the Ansatz $\mathcal U(\theta)$ is updated and new
measurements are performed. Although the general scheme for variational circuits
is pretty simple, lots of details can be deployed regarding the three pieces of
this algorithm.

We propose a model based on the general framework of VQC to tackle the problem
of fitting one or several PDFs flavours using quantum computers. In this case,
the problem to be solved is mathematically reduced to approximate arbitrary
one-dimensional functions within a certain target accuracy. That is, we define
the PDF model to be parametrized by a VQC as
\begin{equation}
  \PDF,
\end{equation}
where $x$ is the momentum fraction of the incoming hadron carried by the given
parton with flavour $i$ (quarks and gluon), so $0 \leq x \leq 1$, at a fixed
initial energy scale $Q_0$. Following this definition, we propose some
superficial modifications to adjust the VQC to this particular problem.

First, we need to introduce the value of $x$ into the circuit. Thus, we modify
the definition of the Ansatz to depend on $\theta$ and $x$, that is $\mathcal
U(\theta) \rightarrow \mathcal U(\theta, x)$. This $x$ value is introduced as
inner circuit parameters following the re-uploading procedure in
Ref.~\cite{perezsalinas:2020reuploading}. The effect of the quantum circuit is
then defined as
\begin{equation}
\mathcal{U}(\theta, x) \ket 0^{\otimes n} = \ket{\psi(\theta, x)}, \label{eq:quantumcircuit}
\end{equation}
which produces a significant change in the output state, since it depends now on
$x$ and not only on $\theta$. The key ingredient in this approach is that, as
the variable $x$ serves as input several times in every circuit, it is possible
to obtain non-linear mathematical structures that allow arbitrary fittings. The
exact design of some $\mathcal U(\theta, x)$ Ansätze are further explained in
Sec.~\ref{sec:ansatze}.

The second ingredient in our model is the way PDF information is extracted from
the quantum circuit. We use the $Z$ Pauli gates to define a series of
Hamiltonians to perform measurements with. Let us consider a $n$-qubit circuit
to run our variational algorithm on. The set of Hamiltonians to build is
\begin{equation}
Z_i = \bigotimes_{j=0}^{n} Z^{\delta_{ij}},
\end{equation}
where $\delta_{ij}$ is the Kronecker delta function.

The choice of this Hamiltonian is heuristic. This model creates as many
Hamiltonians as qubits are available in the circuit, and those Hamiltonians are
created by measuring a certain qubit with the $Z$ Pauli matrix, while all other
qubits remain unmeasured. These observables measures the population of the
states $\ket 0$ and $\ket 1$ of a particular qubit. The hamiltonian is proposed
in order to encode the PDF functions within the probability of measuring a
certain qubit in its excited state. Following the Hamiltonians previously
stated, we can define the function
\begin{equation}
z_i(\theta, x) = \bra{\psi(\theta, x)} Z_i \ket{\psi(\theta, x)}.
\end{equation}
The next step is to relate these $z_i$ functions to the PDF values. We associate
each function $z_i(\theta, x)$ to only one parton $i$. That is, if the model
aims to fit $n$ partons, the circuit width must be $n$ qubits. We define the
qPDF model for flavour $i$ at a given $(x,Q_0)$ as
\begin{equation}
\PDF = \frac{1 - z_i(\theta, x)}{1 + z_i(\theta, x)}.
\end{equation}
With this choice only
positive values are available, although there is no upper bound.
The reason to choose this particular definition is heuristic and is
supported by empirical results detailed in a later section.
It is, however, not a hard constraint, as it is
possible to drop this positivity constraint with a simple re-scaling.
A theoretical motivation can be drawn from the fact that PDF functions can be made non-negative~\cite{Candido:2020yat}
but their values may in principle grow to any real value, see
for instance the gluon PDF in Fig.~\ref{fig:all_flavours}.

\section{Implementation}\label{sec:implementation}

\subsection{Workflow design}

\begin{figure}
  \includegraphics[width=0.4\textwidth]{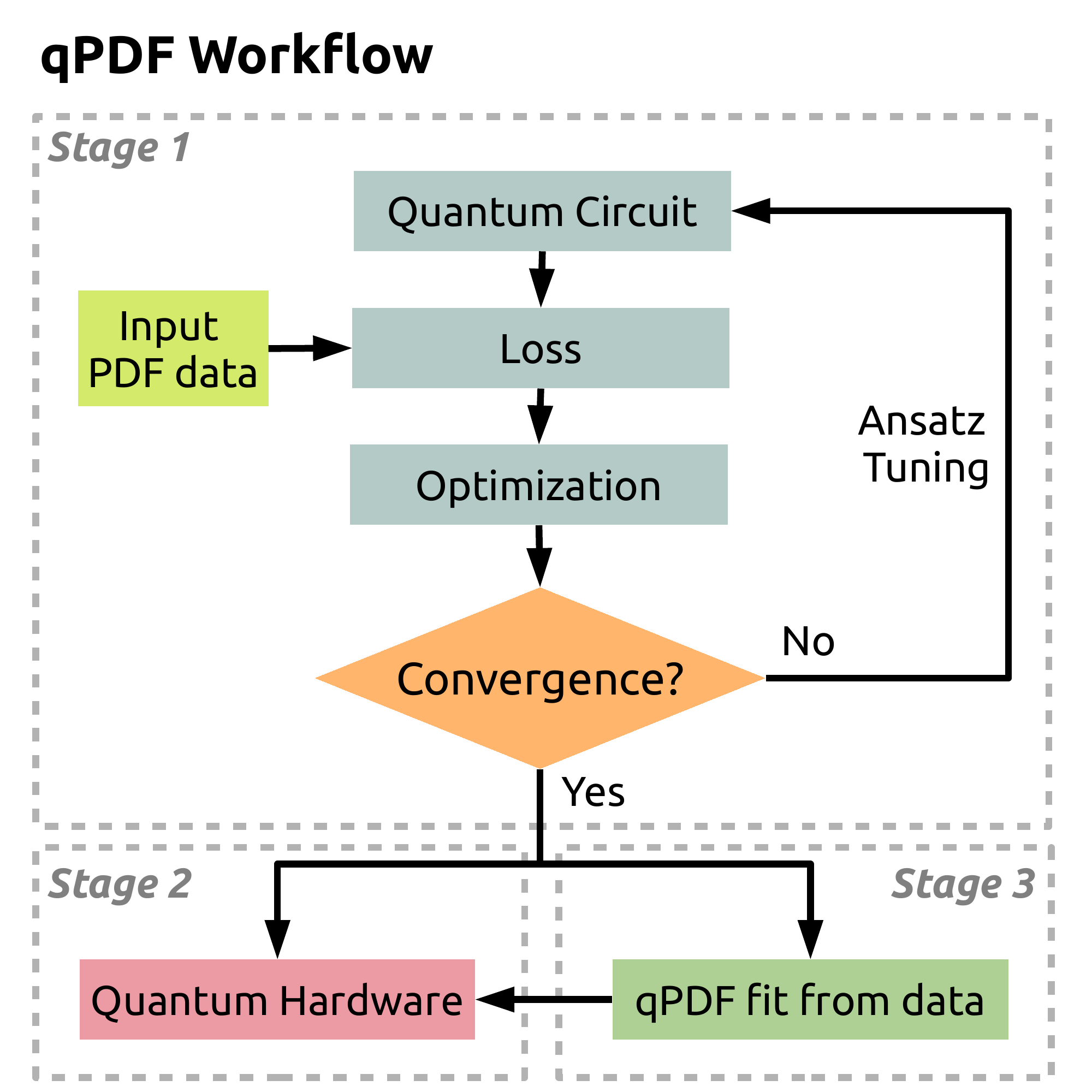}
  \caption{\label{fig:workflow}Schematic workflow for the implementation of
  qPDF.}
\end{figure}

In order to achieve our goal to determine a set of PDFs based on quantum
circuits, we have defined a workflow based on a step-by-step procedure composed
by three stages: (1) the identification of the most adapted quantum circuit
Ansatz for qPDF parametrization, (2) the feasibility study to deploy the qPDF
model into real quantum devices, and finally, (3) the integration of the quantum
circuit model in a global PDF fitting framework.

In Fig.~\ref{fig:workflow} we show schematically the three stages we followed.
Firstly, we perform simulations to identify the best model architecture and
capacity to represent PDF-like functions. This stage is similar to the usual
hyper-optimization tune performed in Machine Learning applications. However, in
our context, we do not have a specific initial Ansatz assumption, thus empirical
tests and fine-tune is required. These simulations are done by computing the
exact wave-function of all quantum states involved in the middle steps of the
algorithm using classical hardware. The expected values for Hamiltonians are
also exactly computed and not measured. The model is then trained to fit PDF
input data generated from the NNPDF3.1 set of PDFs~\cite{Ball:2017nwa}. In the
next section we discuss the details of the procedure and identify the best model
architecture for the qPDF determination.

The second stage studies the possibility to deploy the qPDF model in an actual
quantum device. For this step, we introduce measurements and noise models, and
identify the required number of shots and trials for an acceptable
representation of PDFs.

Finally, as a third and last stage, we use this model in an actual PDF fit based
on experimental data (mainly comprised of LHC measurements). We have integrated the qPDF model
from stage 1 in the NNPDF fitting
framework~\cite{Carrazza:2019mzf,Forte:2020yip}. This implementation opens the
possibility to perform fits on the same datasets of modern PDF releases.

All calculations involving quantum circuits are performed using the quantum
simulation tool {\tt
Qibo}~\cite{efthymiou:2020qibo,stavros_efthymiou_2020_4071702} on classical
hardware. The qPDF model is publicly available through the {\tt Qibo} API. The
experimental implementation of this model was done using {\tt
Qiskit}~\cite{qiskit} from the {\tt OpenQASM}~\cite{cross2017open} code
generated by {\tt Qibo}. The processing of experimental data from the LHC
experiments is done with the \texttt{n3fit}~\cite{Carrazza:2019mzf} code.

\subsection{Ansatz determination}\label{sec:ansatze}

\begin{figure*}[t!]
  \centering
  \subfigure[\hspace{2mm} One layer]{
  \begin{tiny}
    \Qcircuit @R=.75em @C=.4em{
      &\qw & \multigate{7}{U^{l}(\theta_l, \gamma_l, x)} & \qw & \push{\hspace{5mm}} & \qw & \gate{U(\theta_{l,0}, x)} & \ctrl{1} & \qw & \gate{R_z(\gamma_{l,7})} & \qw \\
      &\qw & \ghost{U^{l}(\theta_l, \gamma_l, x)} & \qw & \push{\hspace{5mm}} & \qw & \gate{U(\theta_{l,1}, x)} & \gate{R_z(\gamma_{l,0})} & \ctrl{1} & \qw & \qw & \\
      &\qw & \ghost{U^{l}(\theta_l, \gamma_l, x)} & \qw & \push{\hspace{5mm}} & \qw & \gate{U(\theta_{l,2}, x)} & \ctrl{1} & \gate{R_z(\gamma_{l,4})} & \qw & \qw & \\
      &\qw & \ghost{U^{l}(\theta_l, \gamma_l, x)} & \qw & \push{\hspace{5mm}} & \qw & \gate{U(\theta_{l,3}, x)} & \gate{R_z(\gamma_{l,1})} & \ctrl{1} & \qw & \qw & \\
      &\qw & \ghost{U^{l}(\theta_l, \gamma_l, x)} & \qw & \push{ = \hspace{1.5mm}} & \qw & \gate{U(\theta_{l,4}, x)} & \ctrl{1} & \gate{R_z(\gamma_{l,5})} & \qw & \qw & \\
      &\qw & \ghost{U^{l}(\theta_l, \gamma_l, x)} & \qw & \push{\hspace{5mm}} & \qw & \gate{U(\theta_{l,5}, x)} & \gate{R_z(\gamma_{l,2})} & \ctrl{1} & \qw & \qw &  \\
      &\qw & \ghost{U^{l}(\theta_l, \gamma_l, x)} & \qw & \push{\hspace{5mm}} & \qw & \gate{U(\theta_{l,6}, x)} & \ctrl{1} & \gate{R_z(\gamma_{l,6})} & \qw & \qw &  \\
      &\qw & \ghost{U^{l}(\theta_l, \gamma_l, x)} & \qw & \push{\hspace{5mm}} & \qw & \gate{U(\theta_{l,7}, x)} & \gate{R_z(\gamma_{l,3})} & \qw & \ctrl{-7} & \qw & \\
      & & & & & & & & & & & & & & & & &
      }
    \end{tiny}
  }
  \subfigure[\hspace{2mm} Full Ansatz]{
  \begin{tiny}
    \Qcircuit @R=.46em @C=.4em{
      &\qw & \multigate{7}{\mathcal U(x, \theta, \gamma)} & \qw & \push{\hspace{5mm}} & \qw & \multigate{7}{U^0} & \qw & \push{\cdots} & \multigate{7}{U^l} & \qw & \push{\cdots} & \gate{U^{L-1}_0} & \qw & \meter \\
      &\qw & \ghost{\mathcal U(x, \theta, \gamma)} & \qw & \push{\hspace{5mm}} & \qw & \ghost{U^0} & \qw & \push{\cdots} & \ghost{U^l} & \qw & \push{\cdots} & \gate{U^{L-1}_1} & \qw & \meter \\
      &\qw & \ghost{\mathcal U(x, \theta, \gamma)} & \qw & \push{\hspace{5mm}} & \qw & \ghost{U^0} & \qw & \push{\cdots} & \ghost{U^l} & \qw & \push{\cdots} & \gate{U^{L-1}_2} & \qw & \meter \\
      &\qw & \ghost{\mathcal U(x, \theta, \gamma)} & \qw & \push{\hspace{5mm}} & \qw & \ghost{U^0} & \qw & \push{\cdots} & \ghost{U^l} & \qw & \push{\cdots} & \gate{U^{L-1}_3} & \qw & \meter \\
      &\qw & \ghost{\mathcal U(x, \theta, \gamma)} & \qw & \push{ = \hspace{1.5mm}} & \qw & \ghost{U^0} & \qw & \push{\cdots} & \ghost{U^l} & \qw & \push{\cdots} & \gate{U^{L-1}_4} & \qw & \meter \\
      &\qw & \ghost{\mathcal U(x, \theta, \gamma)} & \qw & \push{\hspace{5mm}} & \qw & \ghost{U^0} & \qw & \push{\cdots} & \ghost{U^l} & \qw & \push{\cdots} & \gate{U^{L-1}_5} & \qw & \meter \\
      &\qw & \ghost{\mathcal U(x, \theta, \gamma)} & \qw & \push{\hspace{5mm}} & \qw & \ghost{U^0} & \qw & \push{\cdots} & \ghost{U^l} & \qw & \push{\cdots} & \gate{U^{L-1}_6} & \qw & \meter \\
      &\qw & \ghost{\mathcal U(x, \theta, \gamma)} & \qw & \push{\hspace{5mm}} & \qw & \ghost{U^0} & \qw & \push{\cdots} & \ghost{U^l} & \qw & \push{\cdots} & \gate{U^{L-1}_7} & \qw & \meter \\
      & & & & & & & & & & & & & & & & &
      }
    \end{tiny}
  } \caption{On the left we show an example of one layer architecture. On the
  right we present the scheme of a full Ansatz circuit including 8 qubits and
  entangling gates. The $U^l(\theta_l, \gamma_l, x)$ from the left figure enters
  the full ansatz as $U^l$. Note that the last layer does not have any
  entangling gate.}
  \label{fig:ansatz}
  \end{figure*}
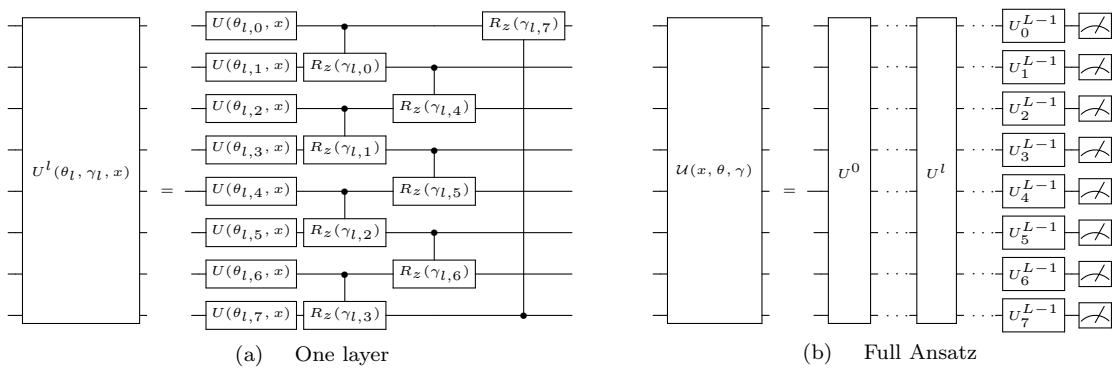

We discuss now the different Ansätze that are considered in this work. Two main
different kinds of Ansätze were designed. The first one, named Weighted Ansatz,
is directly inherited from Ref.~\cite{perezsalinas:2020reuploading}, and
introduces the $x$ variable using the weights and biases scheme, similarly to
Neural Networks. The second one, called Fourier Ansatz, inspired in
Ref.~\cite{schuld:2020encoding}, is related to harmonic analysis and uses
linear and logarithmic scaling to satisfy all values of $x$ involved in PDF
determination, in particular for small and large values of $x$, where
experimental data suffers from larger uncertainties. The main difference between
both Ansätze is the presence or absence of tunable weights.

In the Weighted case, the single-qubit gate serving as building block for the
whole Ansatz is
\begin{equation}
U_w(\alpha, x) = R_z(\alpha_3 \log(x) + \alpha_4) R_y(\alpha_1 x + \alpha_2),
\end{equation}
where $\alpha$ is a four-component set of parameters. Notice that two different axis are
involved in the definition of this gate. This is due to the fact that any two
different Pauli matrices do not commute and leads to the rising of non-linear
mathematical structures, allowing the approximation to be uniformly
accurate~\cite{perezsalinas:2020reuploading, cybenko1989, Hornik1991}. The
presence of both axis allows the possibility to introduce $x$ and $\log(x)$
dependencies to the same gate.

In the Fourier case, we define the gate
\begin{multline}
U_f(\alpha, x) = R_y(\alpha_4)R_z(\alpha_3)R_y(-\pi/2 \log x)\\R_y(\alpha_2)R_z(\alpha_1)R_y(\pi x)
\end{multline}
where the values of the coefficients preceding the $x$ and $\log (x)$ depend on
our dataset. For the specific PDF determination problem presented here, the
values of $x$ are constrained to lie between $10^{-4}$ and $1$, thus the gates
are evaluated at angles between $0$ and $2\pi$.

We use these single-qubit gates to construct layered Ansätze to fit the PDFs.
The reason for this procedure is that we expect to cast more accurately the
output quantum state as more layers are added to the quantum circuit. The layers
have two pieces. First, a layer of as many single-qubit parallel gates as qubits
is applied. Second, a set of entangling gates is added to the circuit. All
entangling gates are controlled $R_z(\gamma)$ gates, where $\gamma$ is also a
tunable parameter. Entangling gates connect one qubit with the next one and then
with the previous one, or viceversa. All layers include the entangling pieces
except for the last one. A scheme depicting the structure of such this circuit
can be viewed in Fig.~\ref{fig:ansatz}. The parameters entering in every gate
are independent for all the other parameters, and all of them are to be
optimized simultaneously. Note that single-qubit circuits cannot have any
entanglement by definition.

For this first tuning stage, we drop the circuit layer with measurement gates and use simulated
final states. The optimization procedure then uses the Pearson's $\chi^2$ loss
function~\cite{pearson:1900chi} to compare the qPDF predictions to the target
central values $f_i$ of NNPDF3.1 NNLO~\cite{Ball:2017nwa}. In this exercise we
always consider a grid of $x$-points distributed between $[10^{-4},1]$
at $Q_0=1.65$ GeV and a maximum of 8 flavours for quarks, antiquarks and the
gluon: $i \in \{\bar{s}, \bar{u}, \bar{d}, g, d, u, s, c(\bar{c})\}$. The
$\chi^2$ covariance matrix is set to a diagonal matrix containing the
$\sigma_{f_i}(x,Q_0)$ uncertainty of the target set.

\begin{table}
\begin{tabular}{
  |c | c!{\vrule width 1.6pt} c | c | }\hline
  \multicolumn{2}{|c!{\vrule width 1.6pt}}{Single-flavour fit} & \multicolumn{2}{c|}{Multi-flavour fit} \\ \hline
  Layers (Parameters) & $\chi^2$ & $\chi^2$ & Layers (Parameters)\\ \noalign{\hrule height .1em}
  1 (32) & \multicolumn{2}{c|}{28.6328} & 1 (32)\\ \hline
  2 (64)& 1.0234 & -- & -- \\ \hline
  3 (96)& 0.0388 & 0.1500  & 2 (72) \\ \hline
  4 (128)& 0.0212 & 0.0320 & 3 (112)\\ \hline
  5 (160)& 0.0158 & 0.0194 & 4 (152)\\ \hline
  6 (192)& 0.0155 & 0.0154  & 5 (192) \\ \hline
\end{tabular}
\caption{Comparison of $\chi^2$ values for the Weighted Ansatz model between the
average of all single-flavour fits (left) and the corresponding multi-flavour
fit (right).}
\label{tab:weighted}
\vspace{0.5cm}
\begin{tabular}{
  |c | c!{\vrule width 1.6pt} c | c | }\hline
  \multicolumn{2}{|c!{\vrule width 1.6pt}}{Single-flavour fit} & \multicolumn{2}{c|}{Multi-flavour fit} \\ \hline
  Layers (Parameters) & $\chi^2$ & $\chi^2$ & Layers (Parameters)\\ \noalign{\hrule height .1em}
 1 (32) & \multicolumn{2}{c|}{900.694} & 1 (32)\\ \hline
  2 (64)& 57.2672 & -- & -- \\ \hline
  3 (96)& 0.0410 & 47.4841  & 2 (72) \\ \hline
  4 (128)& 0.0232 & 0.0371 & 3 (112)\\ \hline
  5 (160)& 0.0165 & 0.0216 & 4 (152)\\ \hline
  6 (192)& 0.0156 & 0.0160  & 5 (192) \\ \hline
\end{tabular}
\caption{Comparison of $\chi^2$ values for the Fourier Ansatz model between the
average of all single-flavour fits (left) and the corresponding multi-flavour
fit (right).}
\label{tab:fourier}
\end{table}

\begin{figure*}[t]
\includegraphics[width=\textwidth]{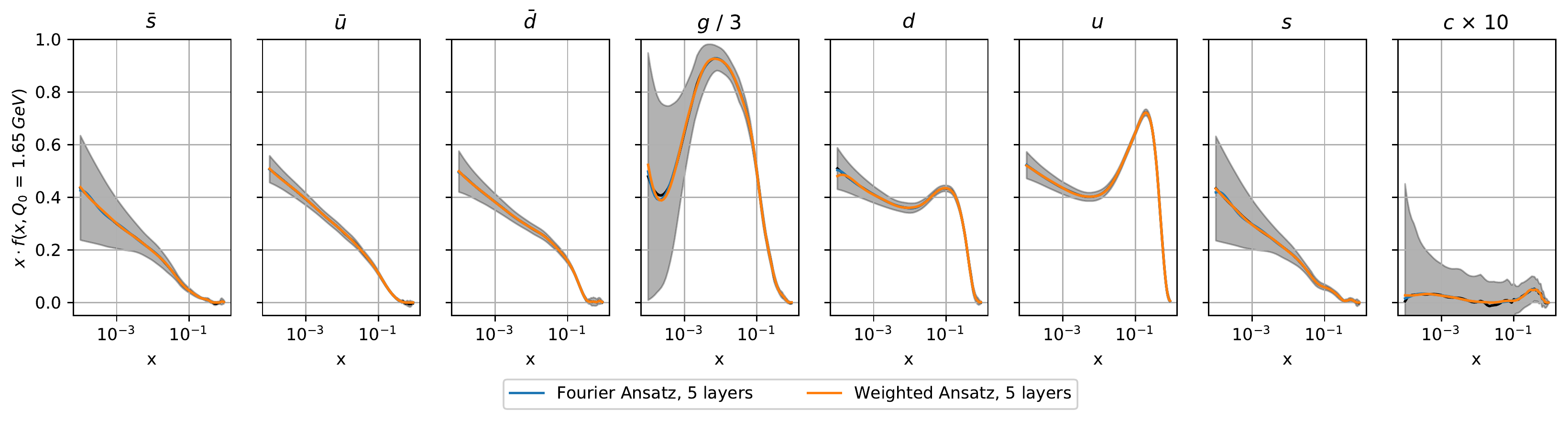}
\caption{Multi-flavour qPDF fits using the Weighted Ansatz (orange curves) and
the Fourier Ansatz (blue curves) with 5 layers and 8
qubits. The mean value and $1\sigma$ uncertainty of the target PDF data is shown
by means of a solid black line and a shaded grey area.}
\label{fig:all_flavours}
\end{figure*}

\begin{figure}
\includegraphics[width=\linewidth]{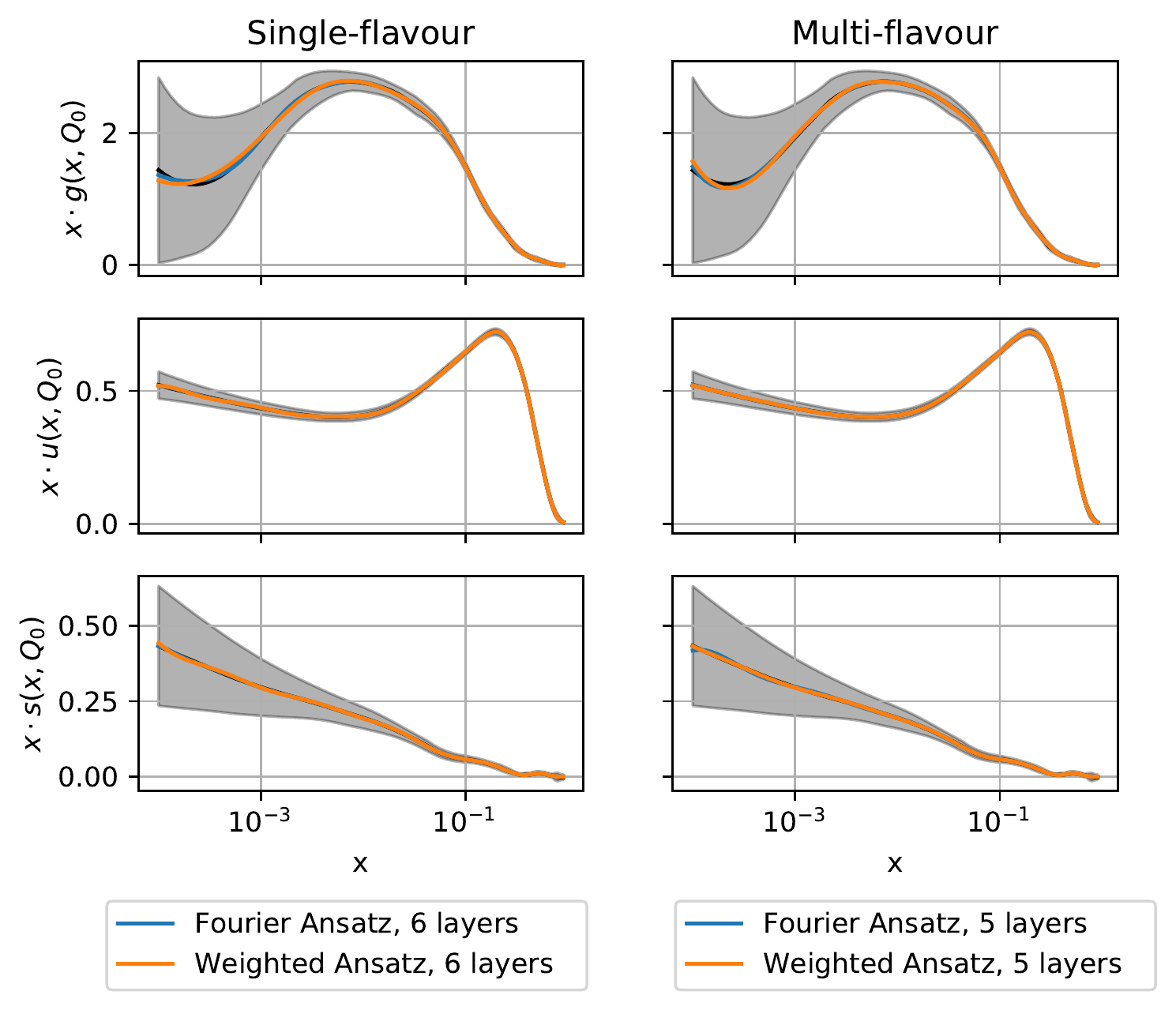}
\caption{Comparison between single-flavour fits (left) and multi-flavour fits
(right) for the gluon, up and strange quarks PDFs. For the single-flavour fits
the Weighted Ansatz (orange curves) and Fourier Ansatz (blue curves) are
composed by 1 qubits and 6 layers. On the other hand for the
multi-flavour fits, the Ansätze are composed by 8 qubits and 5 layers. The mean value and $1\sigma$ uncertainty of the target PDF data is shown
by means of a solid black line and a shaded grey area.}
\label{fig:single_flavours}
\end{figure}

Results summarized in Tables~\ref{tab:weighted} and~\ref{tab:fourier} show the
values for Pearson's $\chi^2$ function both for the Weighted and the Fourier
Ansätze respectively. In both cases, the left column shows an average fit for
all the flavours in a one-by-one fashion, while the right column shows an
optimization for all flavours simultaneously. These table compare the performance
between circuits with similar number of parameters, that is, in every pair
unentangled circuits have a larger number of layers than entangled circuits. The
reason to compare circuits in this way is because entanglement is expected to
improve the overall quality of the fits. The calculations were in this case made
by simulating all the operations on quantum circuits, and the optimization procedure
was done in two steps. First, the {\tt CMA} genetic algorithm is used to find optimal solutions
for single-flavour optimizations \cite{cma}. In the multi-flavour scenario we used
the {\tt L-BFGS-B} function from {\tt scipy}~\cite{l-bfgs, scipy}. The multi-flavour optimizations
start from the corresponding single-flavour results and add the entangling
gates, allowing for a better fitting. In addition, some
results for the final fitting are to be viewed in Figs.~\ref{fig:all_flavours}
and~\ref{fig:single_flavours}.

There are several interpretations that can be claimed from those results. First,
it is clear that entanglement does not suffice to obtain good approximations.
Entanglement can be understood as a quantum resource to extract the correlations
between different qubits, which in this case encode the information of qPDFs
within. On the other hand, every layer of variational gates provides a new step
in non-linearity, which is necessary to represent arbitrary functions. Thus,
entanglement may help to achieve better fittings, as seen in
Tables~\ref{tab:weighted} and~\ref{tab:fourier} for models with the same number
of layers. However, a sufficient number of layers is also mandatory.
Secondly, data unveils the goodness of the Weighted Ansatz with respect to the
Fourier one. Built-in weights grant the model a great representability,
especially in the cases with a small number of layers.

As final Ansatz, we will retain the Weighted one with 5 layers both in the
single-flavour and multi-flavour scenarios. For the sake of
comparison, equivalent Fourier Ansätze
are chosen. In the remainder of this work, we are using the 5-layers
multi-flavour Weighted Ansatz. This circuit has got 5 layers of single-qubit
gates and 4 layers of entangling gates interspersed with the single qubit
layers, up to a total amount of 192 parameters, which is a manageable number.
A detailed comparison in the number of parameters is deployed in Tab. \ref{tab:summary_ansatz}.
The entangling gates are
controlled-$R_z$ gates with one inner parameters, and single-qubit gates are
parameterized through the scheme $w x + b$, where $x$ is the variable for the
PDFs. Logarithmic and linear scales are used together in the same quantum gate.
This configuration is also the first one allowing for a path between all qubits
of the circuit. Results depicted in Table~\ref{tab:weighted} and
Figs.~\ref{fig:all_flavours} endorse the use of this Ansatz. In addition,
tests run on both Ansätze reveiled that the Weighted Ansatz is easier to train using
efficient gradient-based methods such as {\tt L-BFGS-B}.

\begin{table}
    \begin{tabular}{| c | c | c | c | c |}\cline{2-5}
    \multicolumn{1}{c|}{} & \multicolumn{2}{c|}{Single-flavour} & \multicolumn{2}{c|}{Multi-flavour} \\ \cline{2-5}
    \multicolumn{1}{c|}{} & Weighted & Fourier & Weighted & Fourier \\ \hline
    Qubits $(q)$ & \multicolumn{2}{c|}{1 (per flavour)} & \multicolumn{2}{c|}{8} \\ \hline
    Layers $(l)$ & \multicolumn{2}{c|}{5} & \multicolumn{2}{c|}{5} \\ \hline
    \multirow{3}{*}{Parameters} & $2 \cdot l \cdot q$ weights & \multirow{2}{*}{$4 \cdot l \cdot q$} & $16 \cdot l$ weights & \multirow{2}{*}{$32 \cdot l$} \\
     & $2 \cdot l \cdot q$ biases &  & $16 \cdot l$ biases &  \\ \cline{2-5}
     & \multicolumn{2}{c|}{No entanglement} & \multicolumn{2}{c|}{$8 (l - 1)$ entangling} \\ \hline
    \end{tabular}
    \caption{Summary for the Ansätze chosen for this work. The preferred number of l
    ayers was chosen as a compromise between small $\chi^2$ and number of parameters.
    Results depicted in Tables~\ref{tab:weighted} and~\ref{tab:fourier}
    determine that the multi-flavour Weighted Ansatz is our best candidate model.}
    \label{tab:summary_ansatz}
\end{table}

\section{Experimental configuration}\label{sec:quantumhardware}

\begin{figure*}[ht]
  \includegraphics[width=\textwidth]{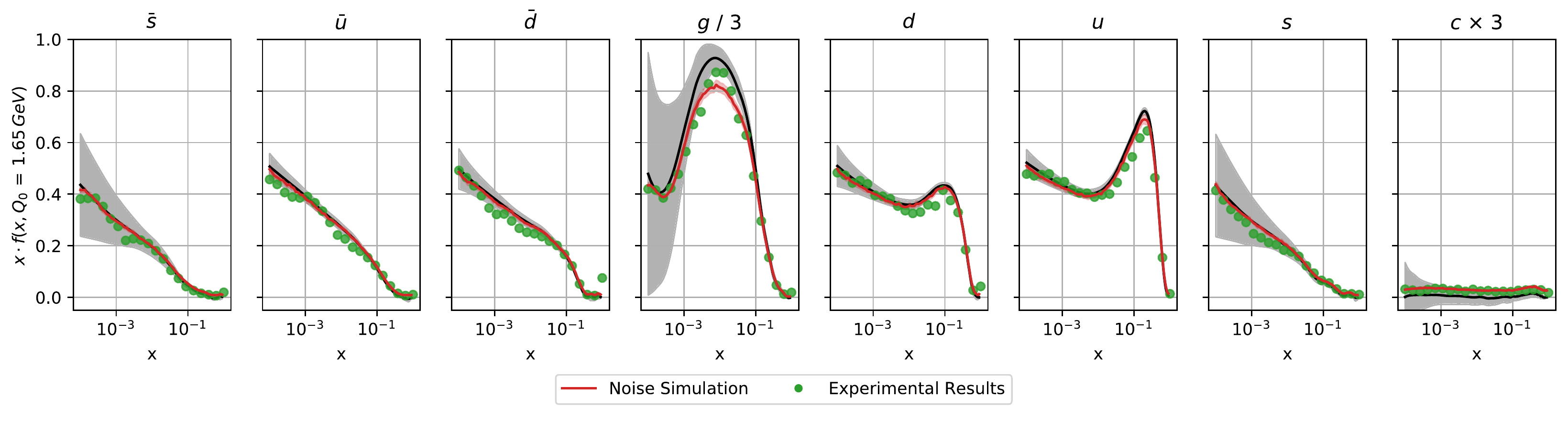}
  \caption{Single-flavour fit for all flavours, using the Weighted Ansatz, for 5
  layers and 8 qubits, that is one qubit per flavour. The red lines represent
  the prediction of the qPDF model with simulated noise from the IBM Athens
  processor \cite{qiskit}. Green points are the results of running the circuit
  on the Athens quantum processor. The mean value and $1\sigma$ uncertainty of the
  target PDF data is shown by means of a solid black line and a shaded grey area.}
  \label{fig:SingleFlavorExperiment}
\end{figure*}

The previous section showed that a low-depth variational Ansatz is capable of expressing the
full set of PDF functions, this section investigates how well that expressibility transfers to a
realistic quantum computer. In order to understand the effects of noise on the
model, the trained single-flavour model was compiled on the IBM Athens quantum processor \cite{qiskit}.
In the single-flavour model, each qubit/parton is fit independently of the others,
and therefore the circuit can be efficiently represented as a rotation of the Bloch sphere.
This fact makes the single-flavour model robust to single-qubit gate errors. Each parton was
evaluated at 20 logarithmically-spaced points between $10^{-4}<x<1$. At each point, the expectation value,  $z_i =
\langle\psi|Z_i|\psi\rangle$, is estimated using $8192$ shots. The evaluation of each point was
repeated five times in order to probe the statistical uncertainty in estimation, it was found that
the estimation was robust to statistical noise. Fig.~\ref{fig:SingleFlavorExperiment} shows the
comparison of running the experiment, and the simulation results. From this figure we deduce that
the single-flavour model produces acceptable results on currently available quantum computers,
and that the {\tt Qiskit} noise simulation environment does a good job of predicting the outcome of
the experiment.

In order to gain an understanding of how the proposed multi-flavour model performs on a
quantum computer, the optimized circuit was simulated with a realistic noise
model. The first step of this simulation is to explicitly include the
measurement gates in the circuit as can be seen in Fig.~\ref{fig:ansatz}.
Each qubit represents a particular parton, and therefore the qubits should be
measured independently. The goal of the measurement is to estimate $z_i =
\langle\psi|Z_i|\psi\rangle$, this is achieved in a given number of shots by
subtracting the number of occurrences of measuring 1 from the number of
occurrences of measuring 0, and normalizing by number of shots.

\begin{figure*}[ht]
  \includegraphics[width=\textwidth]{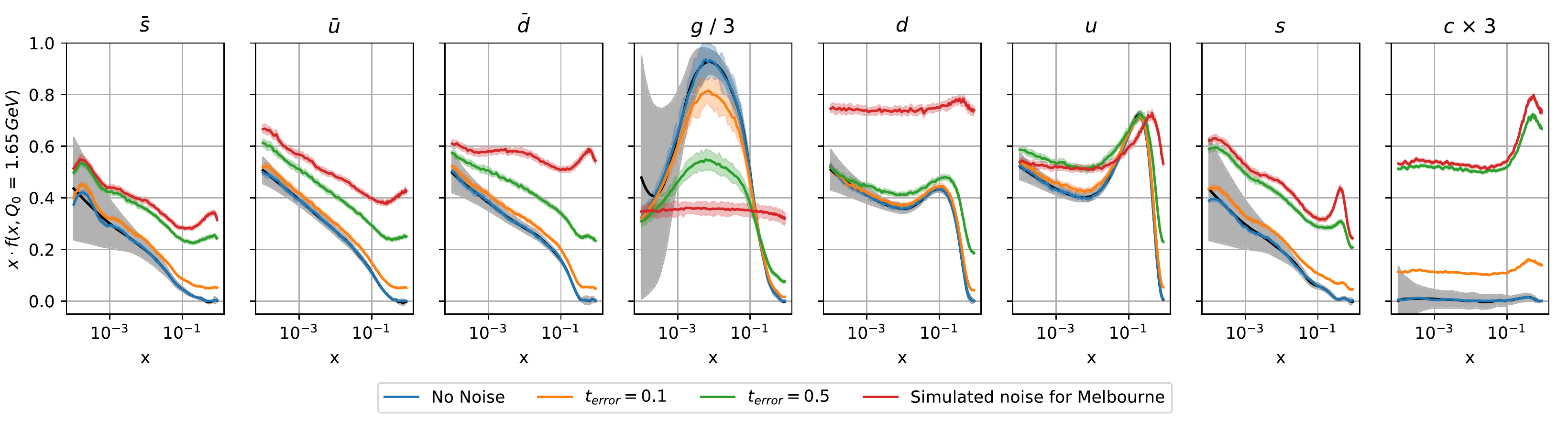}
  \caption{Multi-flavour fit for all flavours, using the Weighted
    Ansatz, for 5 layers and 8 qubits, that is one qubit per
    flavour. Blue lines are the mean and the blue shadowed area the
    $1\sigma$ uncertainty of the circuit measurement results for an
    ideal noise free quantum device. The red curve refers to simulated
    circuit measurements using the noise model for the IBM Melbourne
    processor \cite{qiskit}. Similarly, green and orange curves
    show simulation results with noise reduced by 50\% and 90\%
    respectively. The mean value and $1\sigma$ uncertainty of the
    target PDF data is shown by means of a solid black line and a
    shaded grey area.}
  \label{fig:Ideal_8000}
\end{figure*}

The ability of the circuit to reproduce the PDF was first simulated on an ideal
quantum computer using {\tt{Qiskit}} \cite{qiskit}. The simulation was performed
with $8192$ shots as this value corresponds to the maximum number of
shots permitted per run on IBM quantum processor. It was found that this number
of shots was more than sufficient to converge the estimate of
$\langle\psi|Z|\psi\rangle$, and thus accurately reconstruct the PDF. This is
shown in Fig.~\ref{fig:Ideal_8000}.

In order to simulate the effect of a realistic noise model, the IBM Melbourne
quantum processor was chosen \cite{qiskit}. The Melbourne processor is the only
device that is publicly available through the IBM Quantum Experience that has
enough qubits to fit the optimized circuit. The 8 qubit optimized circuit was
mapped onto Melbourne in such a way to minimize the $\chi^2$.

The errors on the Melbourne device were found to drastically deteriorate the
estimation of the PDF as can be seen in Fig.~\ref{fig:Ideal_8000}. This
analysis has shown that while it is possible in theory to fit a PDF using a
quantum computer, the noise in the current state-of-the-art quantum processors
is still too high to reconstruct the PDF accurately.

\begin{figure}[h]
\centering
  \includegraphics[width=0.4\textwidth]{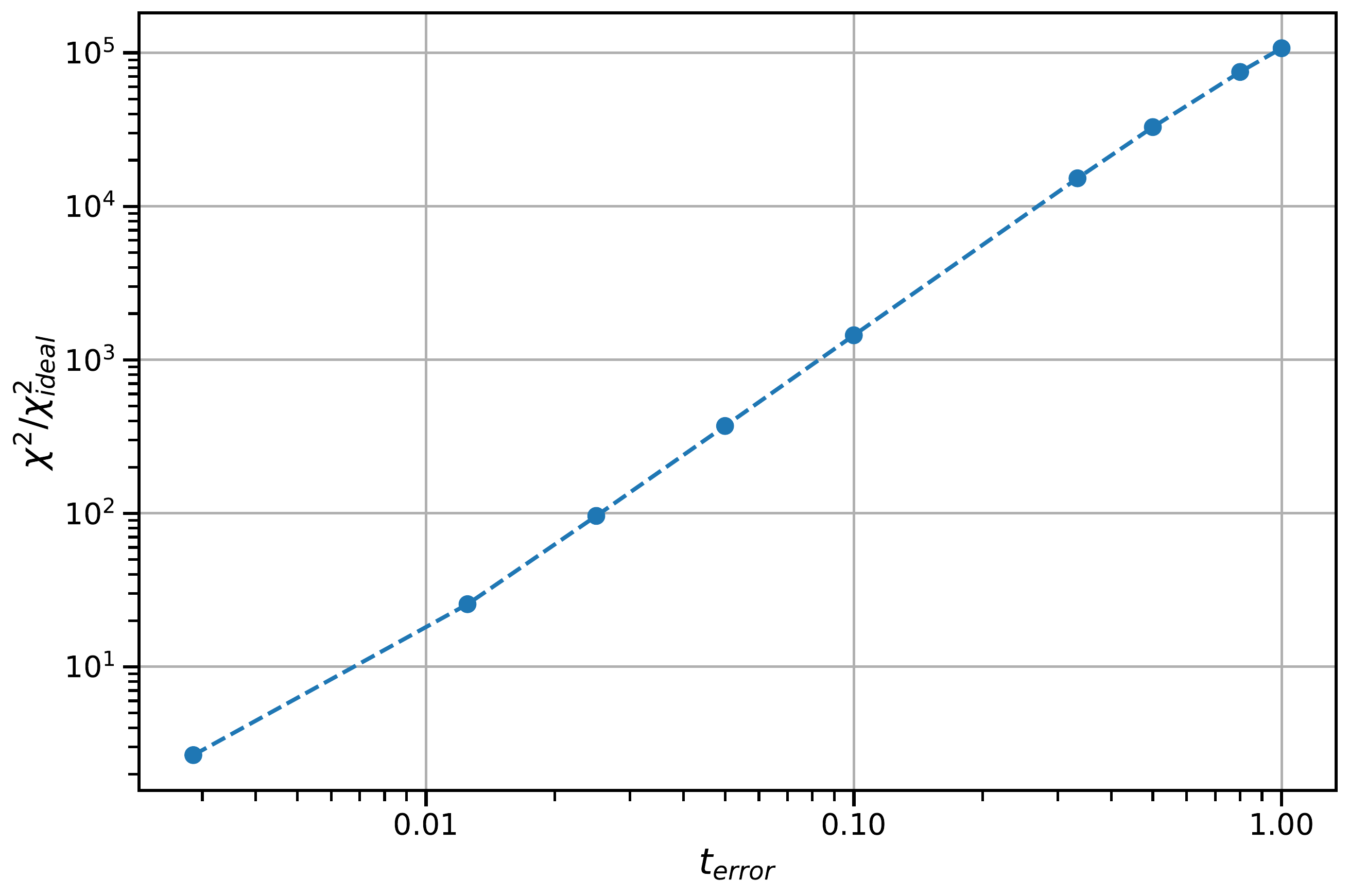}
  \caption{The error as a function of the error interpolation parameter $t_{\text{error}}$. The $y$-axis is given as the ratio between the error, $\chi^2$, and the error on an ideal quantum computer, $\chi_{\text{ideal}}^2$.}
  \label{fig:errorScaling}
\end{figure}

Another question that can be asked is, how robust must the quantum device be in
order to have an acceptable representation of the PDF? To answer this question,
a simplified version of the Melbourne device was created. In this simplified
Melbourne, all the qubits and connections were taken to have identical noise
characteristics, specifically all single gate, double gate, and readout errors
were set to the best values from the real Melbourne processor. With this
simplified device, the noise models can be uniformly scaled down to interpolate
between an ideal quantum computer and a Melbourne-like device using a parameter
$t_{\text{error}}$, where $t_{\text{error}}=0$ corresponds to an ideal quantum
computer, and $t_{\text{error}}=1$ corresponds to the simplified Melbourne
device. Fig.~\ref{fig:errorScaling} shows what happens to the cost function
$\chi^2$ as $t_{\text{error}}$ is varied.

\section{PDF determination from experimental data}\label{sec:lhc}

In the previous sections we have described the process of finding a final Ansatz
which can encode the full complexity of the physical PDFs by training to
already known results. Furthermore, we have verified the possibility to deploy
such model on real quantum devices. These steps correspond with stages 1 and 2
of our workflow (Fig.~\ref{fig:workflow}) where the PDF is treated as a known
quantity. In reality, however, the only data that one has access to are the
experimental measurements of physical observables performed at experiments (for
instance, physical cross sections measured at the LHC).

The next stage of this work is to prove that this methodology can also replace the neural networks at the core of the
NNPDF methodology for fitting PDFs.
Although still far from being of practical usaged (see for instance Fig. \ref{fig:all_flavours})
we show, in the simulator, that a hybrid VQE could indeed replace neural networks as an universal function approximator
for complex problems such as the one posed by parton distribution functions.

In this final section we start by describing the NNPDF methodology and what changes are needed to its latest implementation
(described in~\cite{Carrazza:2019mzf}) to perform a full fit.
We use the NNPDF3.1 dataset which includes deep-inelastic scattering (DIS) and hadronic collider data.
We end with a comparison between our resulting PDF (qPDF) and the latest NNPDF release (NNPDF3.1) and prove that the
results are perfectly usable in an actual computation of physical observables.

\subsection{The NNPDF fitting methodology}
The two main aspect that define the NNPDF methodology are the Monte Carlo approach to the uncertainties of
experimental measurements and the usage of Neural Networks (hence the name) to model the PDFs.
In this section we outline some of the most relevant aspects of the NNPDF methodology, for a more in-depth review please consult~\cite{Ball:2014uwa}.

The first step of the methodology is the generation of ``data replicas''.
This procedure propagates the experimental uncertainties into the PDF fit by leveraging the covariance matrices provided
by the experiments by creating between 100 and 1000 artificial copies of the data as if they were produced by
independent measurements.

The full PDF fitted in this methodology follows the functional form for each parton $i$:
\begin{equation}
    f_i(x,Q_0) = x^{-\alpha_i} (1-x)^{\beta_i} {\rm NN}_i(x), \label{eq:PDFdefinition}
\end{equation}
where the fitted NN is prepended by a preprocessing factor per parton $x^{-\alpha}(1-x)^{\beta}$.
This factor ensures the correct behaviour at very small (close to 0) and very large (close to 1) values
of $x$, where there might not be enough experimental data to properly constraint the NN.
This function constrains all free parameters that define the behaviour of the PDF.
The functions defined in Eq.~\eqref{eq:PDFdefinition} however cannot be directly compared to experimental data, instead one would have to convolute them
with the partonic cross section in order to obtain a physical prediction that can be compared to the result of an
experiment,
\begin{equation}
    P = \displaystyle\int \dd{x_{1}}\dd{x_{2}}  f^{i}_{1}(x_{1}, q^{2})f^{j}_{2}(x_{2}, q^{2})|M_{ij}(\{p_{n}\})|^{2}, \label{eq:convolution}
\end{equation}
where $x_{1}, x_{2}$ are the momentum fraction carried by the two colliding partons and the indices $i$ and $j$ run over
all possible partons.
$M_{ij}$ is the matrix element for the given processes and $\{p_{n}\}$ represents the phase space
for a $n$-particles final state.
Performing this integral numerically per training step, per experimental data point, would be completely impracticable.
Instead the theoretical predictions are approximated as a product between the PDF model and a fastkernel table (FK table) encoding
all the relevant information on the computation as described in Refs.~\cite{Ball:2010de,Bertone:2016lga}.

The optimization of the function defined in Eq.~\eqref{eq:PDFdefinition} consists then in the minimization of a $\chi^{2}$ defined as:
\begin{equation}
    \chi^2 = \sum_{i,j}^{N_{\rm dat}} (D-P)_i \sigma_{ij}^{-1} (D-P)_j, \label{eq:chi2}
\end{equation}
where $D_{i}$ and $P_{i}$ are respectively the $i$-nth data point from the training set and
its theoretical prediction and $\sigma_{ij}$ is the experimental covariance matrix provided by the experimental
collaborations.

This procedure is then repeated for each of the artificial replicas.
Note that the theoretical predictions are always the same, so the only change between replicas is in the experimental
data points.
The final central value for the PDF is then the average over all replicas, while the error bands are given by
taking the envelope that contains 68\% of all replicas.

\subsection{\texttt{Qibo}-based \texttt{n3fit}}

The latest implementation of the latest iteration of the NNPDF methodology is described in Ref.~\cite{Carrazza:2019mzf}.
This implementation is very modular and one can seamlessly swap the \texttt{Tensorflow} based backend by any other
provider.
{\tt Qibo}, which is also partially based on \texttt{Tensorflow} can be easily integrated with the NNPDF methodology.

Note that all results in this section corresponds to the simulation of the quantum device on classical hardware.
Such a simulation is very costly from a computational point of view which introduces a number of limitations
that need to be addressed in order to produce results in reasonable time frames.

{\bf FK reduction}: the definition of the quantum circuit depends on both the set of parameters $\theta$ and the value of
the parton momentum fraction $x$ (see Eq.~\eqref{eq:quantumcircuit}) which means the circuits needs to be simulated
once per value of $x$.
The union of all FK tables for all physical observables (following Eq.~\eqref{eq:convolution}) amounts to
several thousand values of $x$.
Since such a large number of evaluations of the quantum circuit is impracticable, we introduce a further approximation
where each partial FK table is mapped to a fixed set of 200 nodes in the x-grid.
This simplification introduces an error to the total $\chi^{2}$ of the order of $\Delta\chi^{2} = 0.14 \pm 0.01 $ when averaged
over PDF members.
This error on the cost function is however negligible for the accuracy reached in this work.

{\bf Positivity}: in the fitting basis, as defined in section~\ref{sec:qcpdf}, the PDF cannot go negative.
Physical predictions however are computed in the flavour basis~\cite{Ball:2008by} where the rotation between basis
can make some results go negative.
However, physical observables (differential or total cross sections) cannot be. This physical constraint is included in NNPDF3.1 via fake pathological datasets.
These have not been implemented for qPDF as they correspond to a fine-tuning of the methodology which is beyond the
scope of this work.

\begin{figure}
    \centering
    \includegraphics[width=0.5\textwidth]{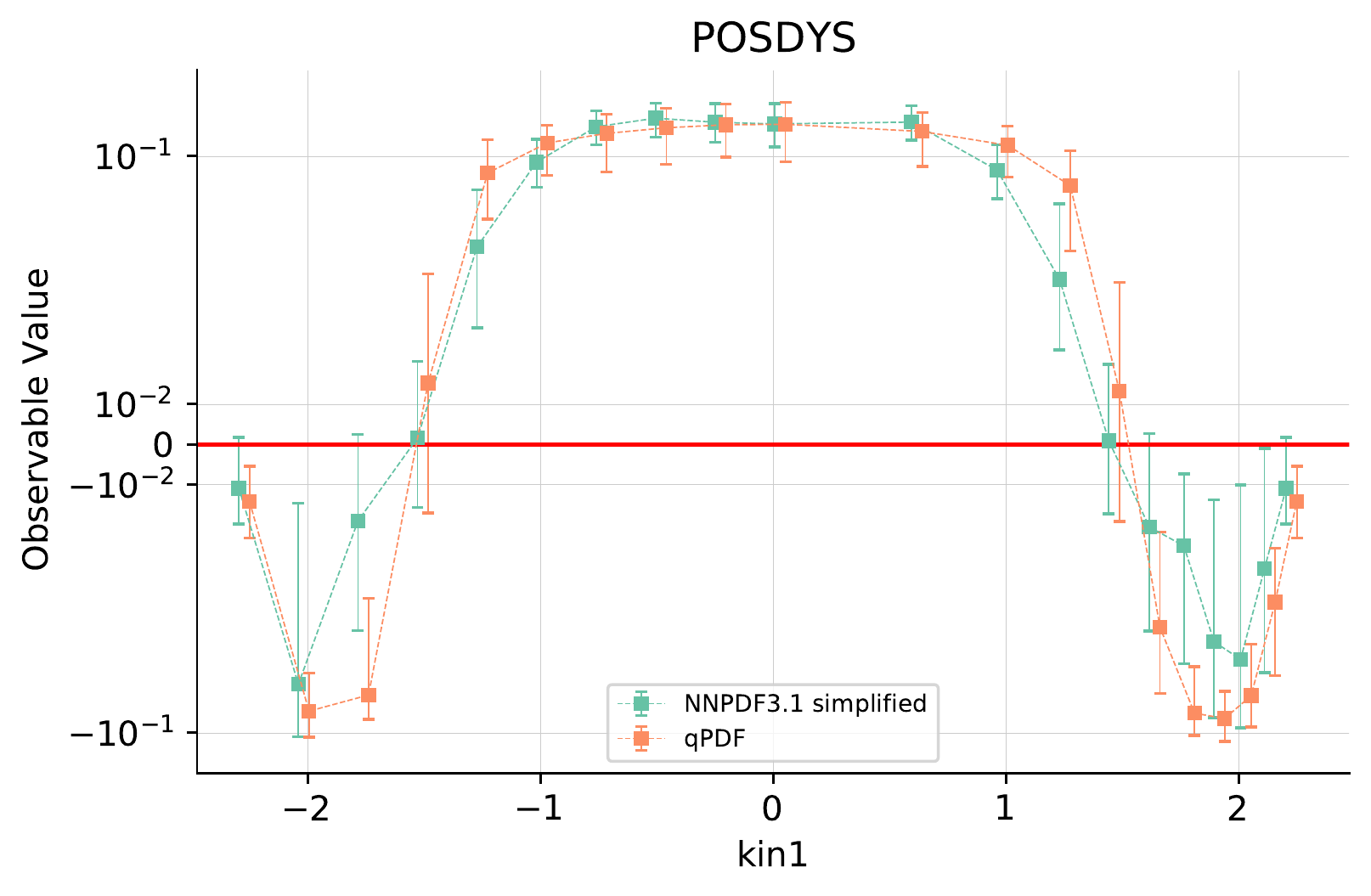}
    \caption{Predictions for a toy $s\bar{s}$ initiated Drell-Yan process with qPDF and a simplified version of NNPDF3.1
    where the positivity constraint has been removed.}\label{fig:positivity}
\end{figure}

The removal of the positivity constraint from the fit introduces an unphysical distortion to the results as the PDF
could produce negative predictions for physical predictions. Such results are unphysical because they would correspond
to situations in which the probability of finding a particular phase space configuration is negative, which makes no
sense.
In Fig.~\ref{fig:positivity} we compare the ``negativity'' between qPDF and a version of NNPDF3.1 with the positivity constraints removed.
We observe that both fits behave similarly, proving such unphysical results are a consequence of the removal of the
constraint rather than a problem in the qPDF methodology.

{\bf Momentum Sum Rule}: the PDFs as defined in Eq.~\ref{eq:PDFdefinition} are normalized such that~\cite{Ball:2014uwa},
\begin{equation}
    \frac{\displaystyle\int_0^1 dx \; x\,f_{g}(x,Q_0)}{1-\displaystyle\int_0^1 dx  x f_{\Sigma}(x,Q_0)} \simeq 1,
\end{equation}
this equation is known as the momentum sum rule and it is imposed in \texttt{n3fit}
through an integration over the whole range of x which is impracticable
in this implementation for the reasons mentioned above.
Instead, in qPDF these are only checked afterwards, finding a good agreement with the expected values (despite not being
imposed at fitting time).
Indeed, for qPDF the result for the average over all replicas is:
\begin{equation}
    \frac{\displaystyle\int_0^1 dx \; x\,f_{g}(x,Q_0)}{1-\displaystyle\int_0^1 dx  x f_{\Sigma}(x,Q_0)} = 1.01\pm 0.01.
\end{equation}
which is to be compared with the NNPDF3.1 result of $1.000 \pm 0.001$, where the constraint was imposed at fit time.

\subsection{qPDF}
Once all ingredients are implemented, we are in a position to be able to run a NNPDF3.1-like fit using the new
prescription based on the VQE and the \texttt{Qibo} library.
As a base reference for the comparison we take the NNPDF3.1 NNLO fit~\cite{Ball:2017nwa},
which is the latest release by the NNPDF collaboration.
The plots comparing the NNPDF sets with qPDF are then produced using a
\texttt{reportengine}~\cite{zahari_kassabov_2019_2571601} based internal NNPDF tool.

The dataset included in this fit correspond to that of NNPDF3.1, which is detailed in
Section 2.1 of~\cite{Ball:2017nwa} and
includes data from deep-inelastic scattering experiments, fixed-target Drell-Yan-like data
and hadronic collider data from experiments at Tevatron and LHC.

\begin{figure}
    \centering
    \includegraphics[width=0.5\textwidth]{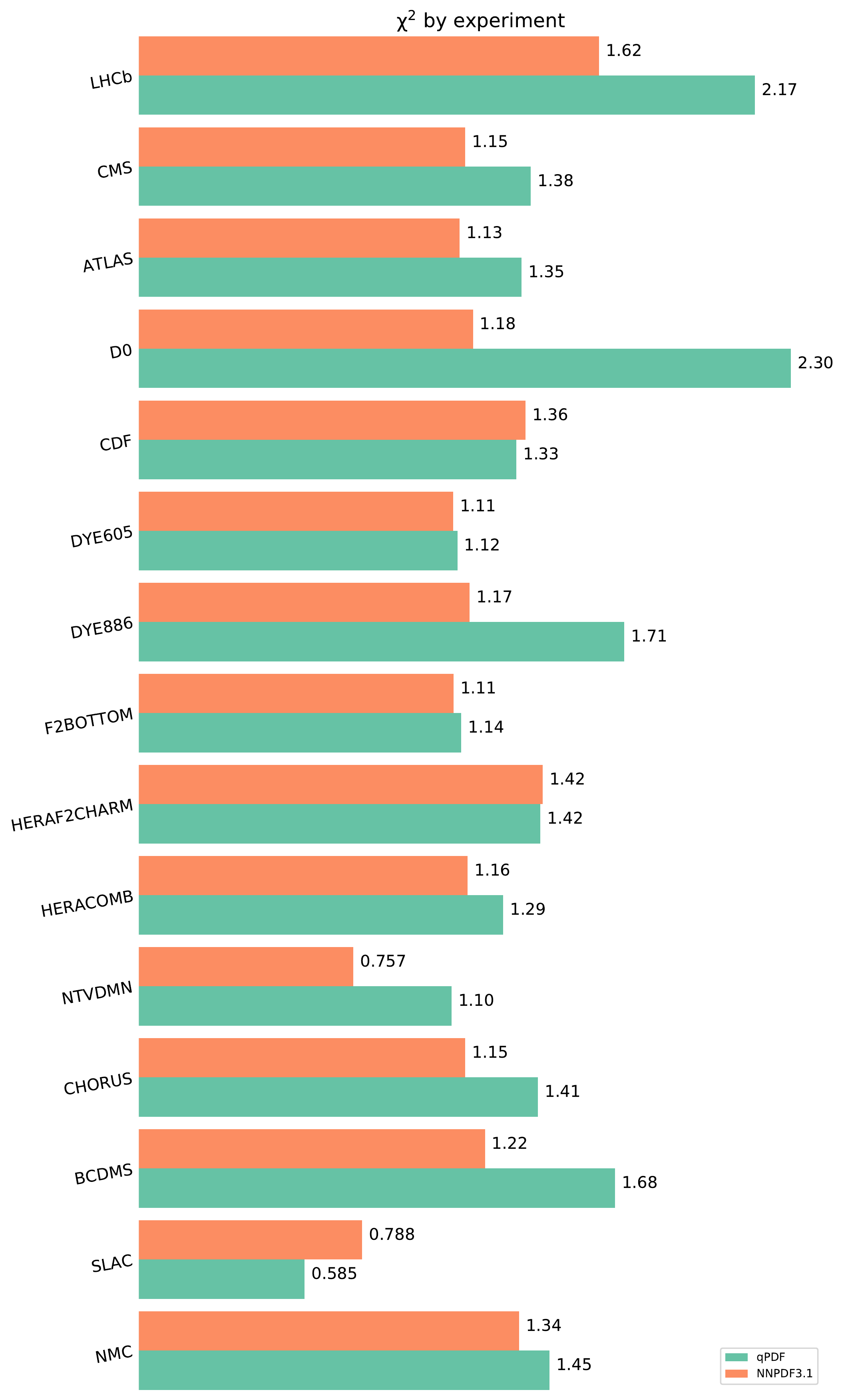}
    \caption{$\chi^{2}/N$ per experiment grouping. There is a deterioration of the goodness of the
        fit (measured by the $\chi^{2}$) for some of the experiments for the central value.
        The goodness of the fit is very similar between the reference and
        qPDF for most of the experiments being considered.}\label{fig:expchi2}
\end{figure}

\begin{figure}
    \centering
    \includegraphics[width=0.5\textwidth]{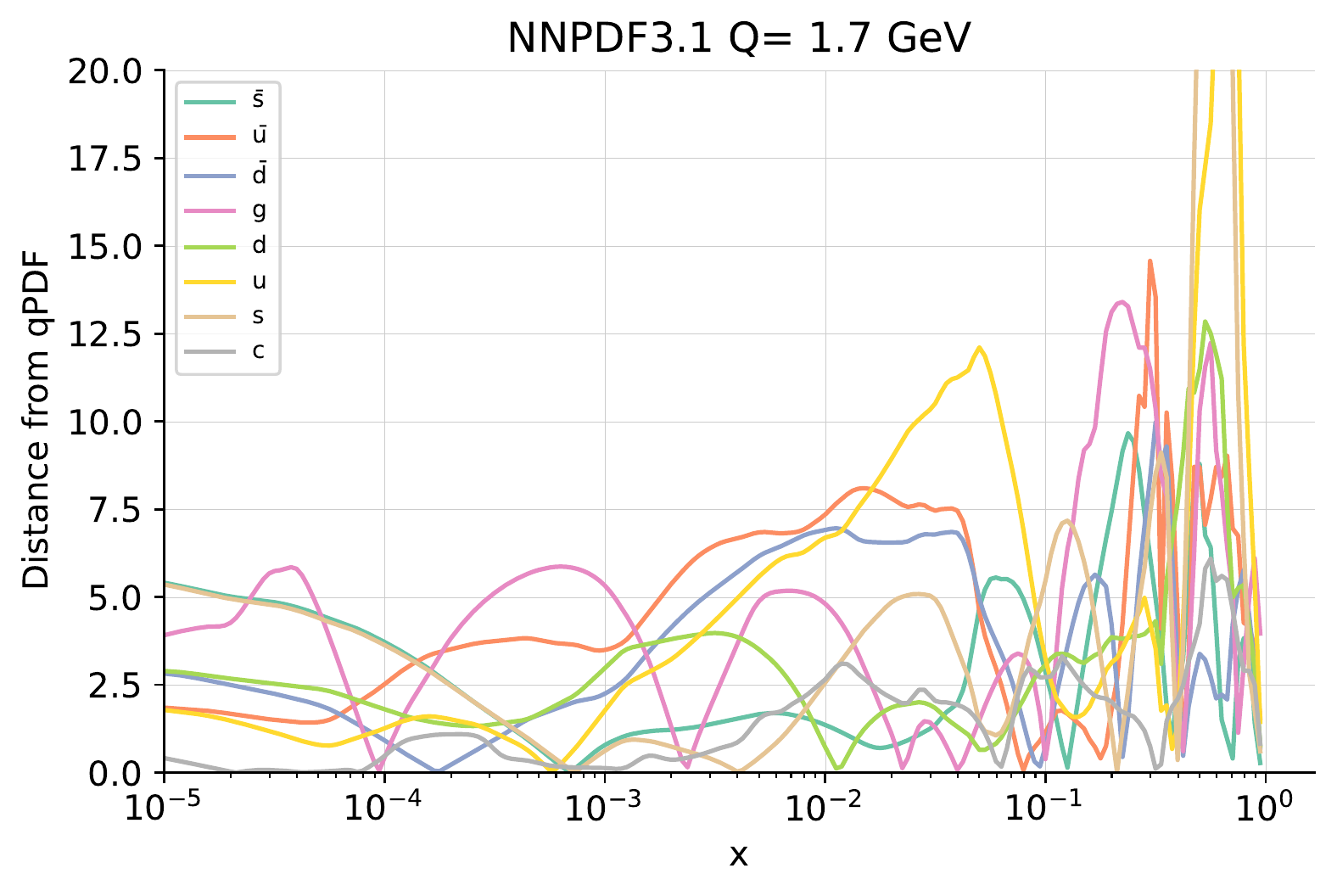}
    \caption{Distance (as defined by Eq.~\eqref{eq:distance}) between qPDF and NNPDF3.1. When the distance is kept
    under $d(f_{i}, r_{i})=10$ the two fits are 1-$\sigma$ compatible. All partons except for $u$ and $s$ are below or
    around the 1-$\sigma$ distance for the entire range considered. Note however, by comparing to
Fig.~\ref{fig:all_flavours} that the fits for both the $u$ and $s$ quarks are compatible in the most relevant regions
for these particles.}\label{fig:distance}
\end{figure}

We can start by comparing the $\chi^{2}/N$ result for the datasets that have been considered in the fit,
shown in Fig.~\ref{fig:expchi2}.
One would expect a perfect fit when $\chi^{2}/N = 1$,
however this is not the case even in the reference and it is due to a combination of missing higher order corrections
(a lack of a better theory) or inconsistencies in the experimental results,

The similarity on the phenomenological results obtained by both fitting methodologies as shown in Fig.~\ref{fig:expchi2}
is well understood as well by looking at the distance plots between the qPDF~and the reference in
Fig.~\ref{fig:distance},
\begin{equation}
    d^2(f_{i}, r_{i}) = \frac{\langle f_{i} \rangle - \langle r_{i} \rangle}{\frac{1}{N_f}\sigma(f_{i})^2 +
    \frac{1}{N_{r}}\sigma(r_{i})^2},\label{eq:distance}
\end{equation}
where $i$ is the flavour being considered and $f$ and $r$ corresponds to qPDF~and the reference (NNPDF3.1)
respectively. The central value is taken over the $N$ replicas of the set, generally of the order of 100.

Indeed, for most partons the difference between both fits are under the 1-$\sigma$ level (distance equal to 10 for 100
replicas) growing up to 2-$\sigma$ for the $u$ and $s$ quarks.

\begin{figure*}
    \centering
    \subfigure[\ Gluon pdf.]{
        \includegraphics[width=0.31\textwidth]{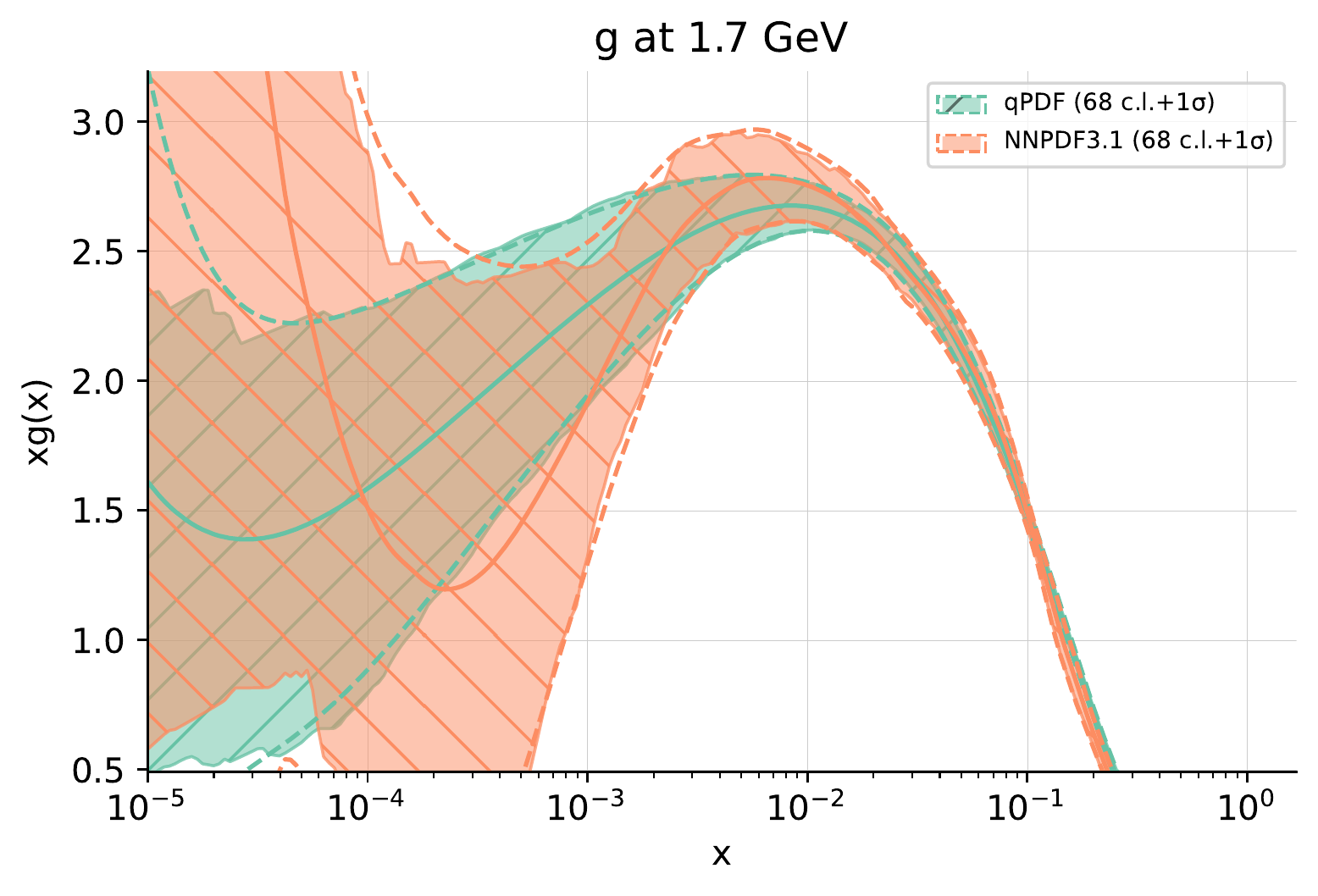}
    }
    \subfigure[\ u quark pdf.]{
        \includegraphics[width=0.31\textwidth]{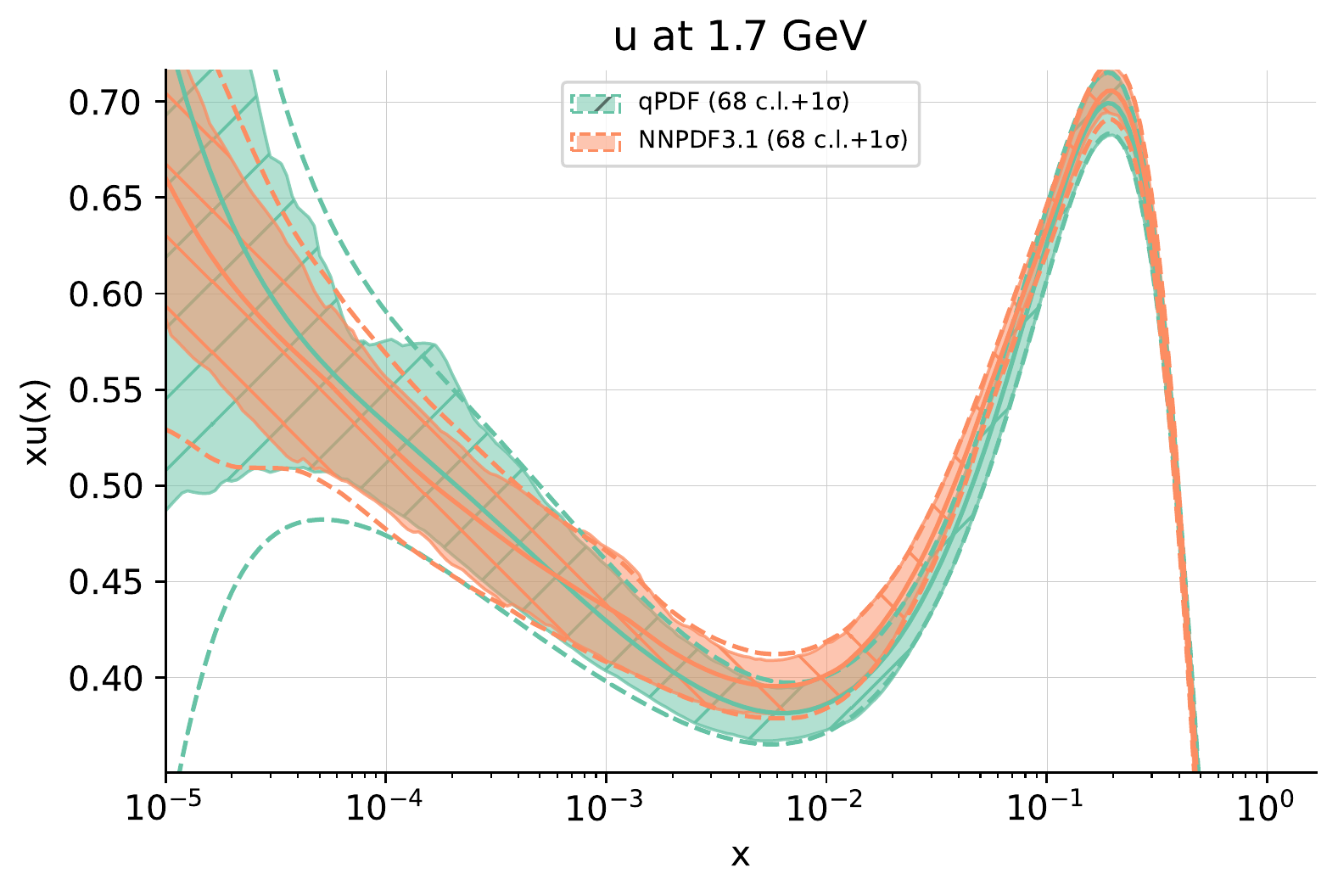}
    }
    \subfigure[\ s quark pdf.]{
        \includegraphics[width=0.31\textwidth]{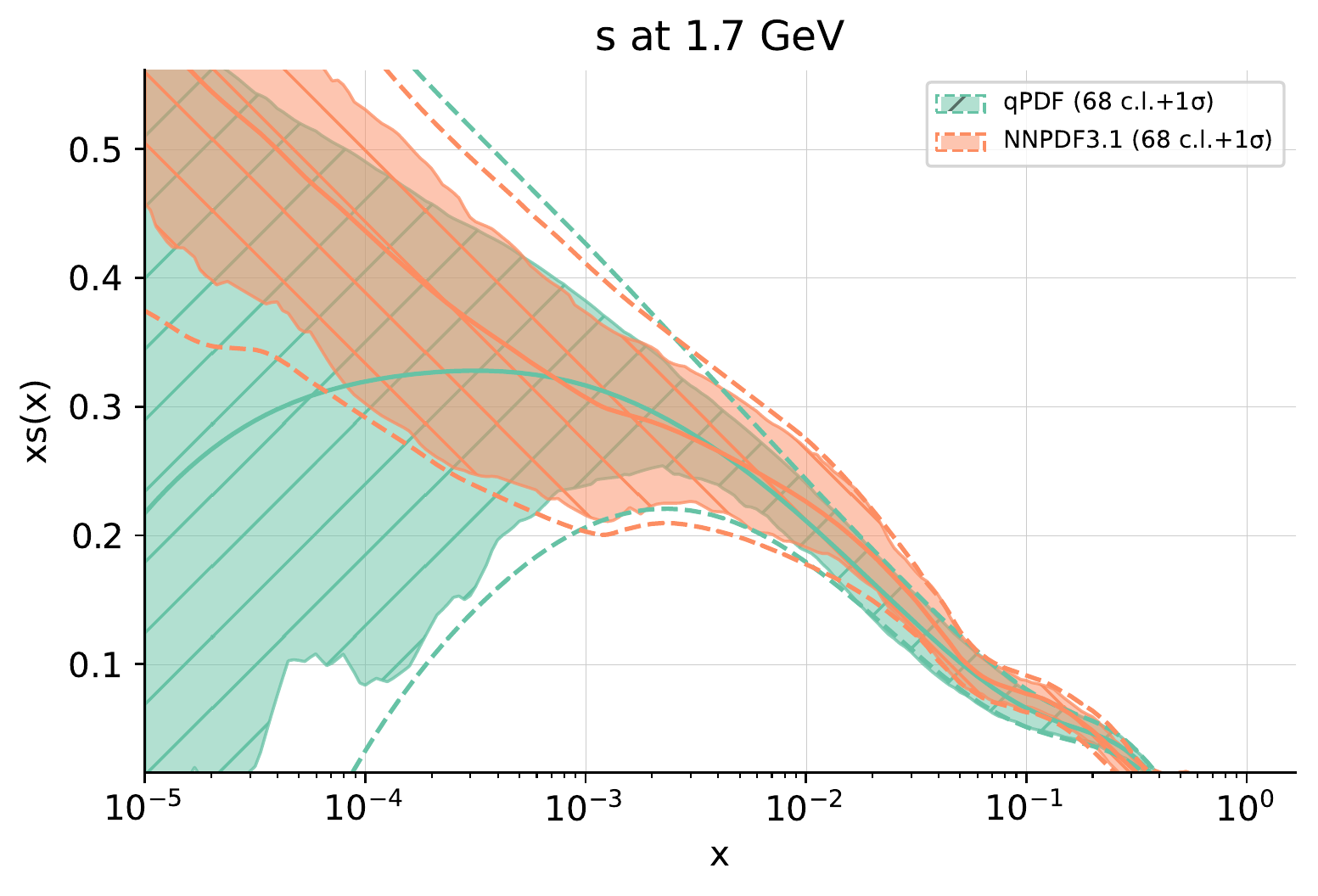}
    }
    \caption{Fit results for the gluon and the $u$ and $s$ quarks. As previously seen in
    Fig.~\ref{fig:all_flavours}, qPDF is able to reproduce the features of NNPDF3.1. We now see this is also true when
the fit performed by comparing to data and not by comparing directly to the goal function. The differences seen at
low-x can be attributed to the lack of data in that region.}\label{fig:fitperflavour}
\end{figure*}

This point is clearly seen in Fig.~\ref{fig:fitperflavour} where we compare the
published PDFs (with their corresponding error bars) for the gluon and the $d$ and
$u$ quarks. We note that for these quark flavours the qPDF central result is almost
always within the 1-$\sigma$ range of the reference, with an overlapping error
band for the whole considered range.

\begin{figure*}
    \centering
    \subfigure[\ Atlas jets data differential in rapidity~\cite{Aad:2014vwa}.]{
        \includegraphics[width=0.31\textwidth]{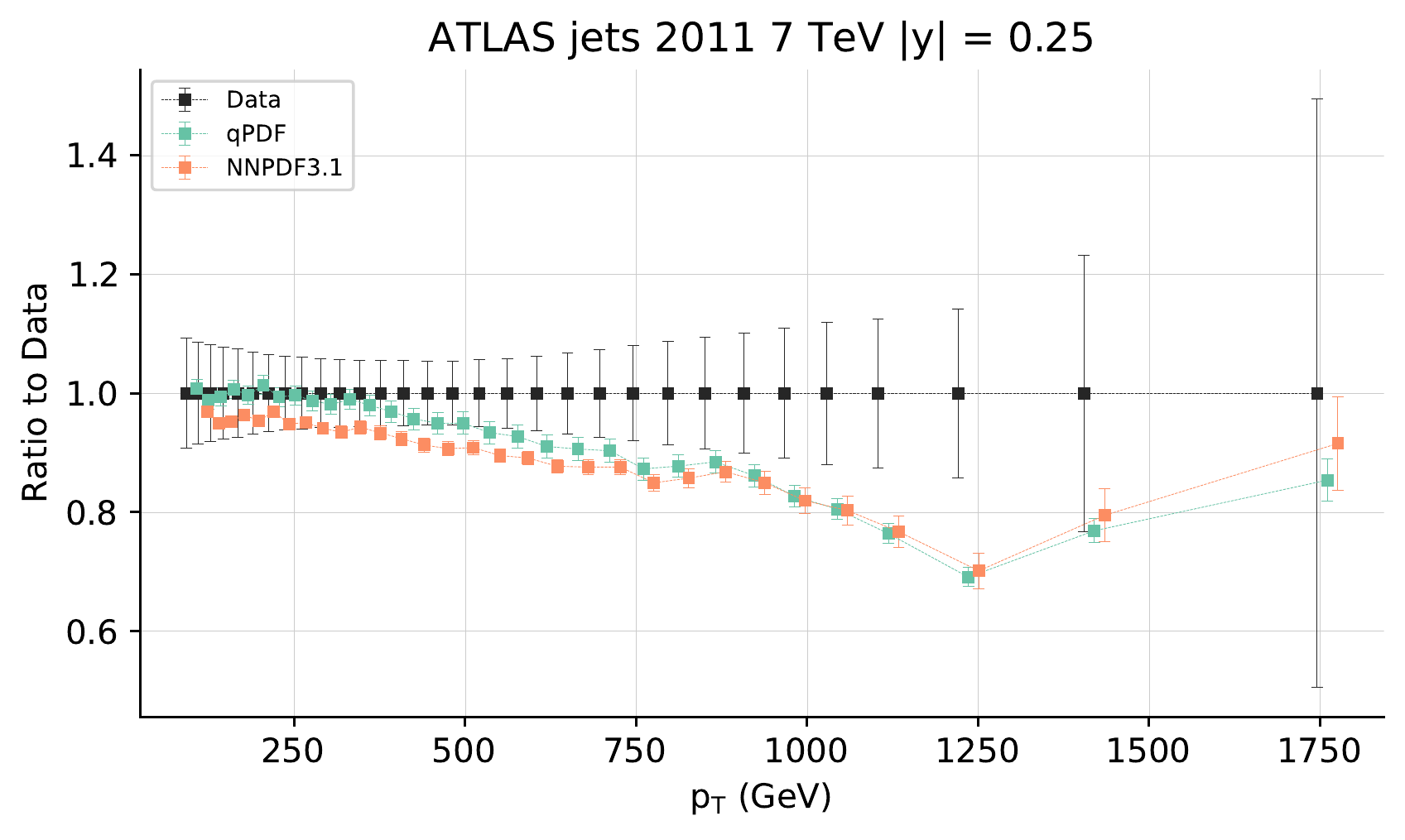}
    }
    \subfigure[\ CMS Z differential in rapidity for fixed value of $p_{T}$~\cite{Khachatryan:2015oaa}.]{
        \includegraphics[width=0.31\textwidth]{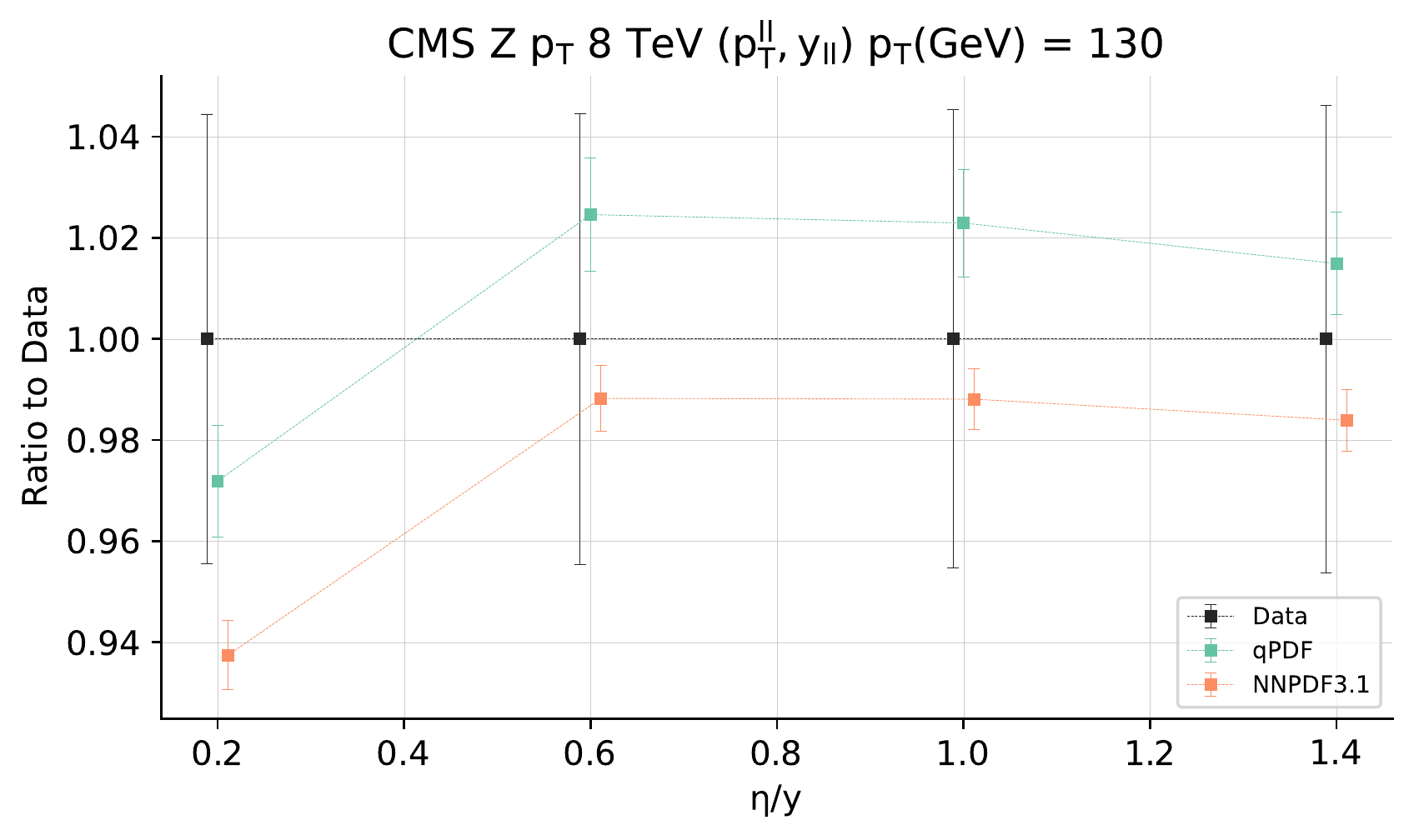}
    }
    \subfigure[\ LHCb, Z cross section differential in rapidity~\cite{Aaij:2012vn}.]{
        \includegraphics[width=0.31\textwidth]{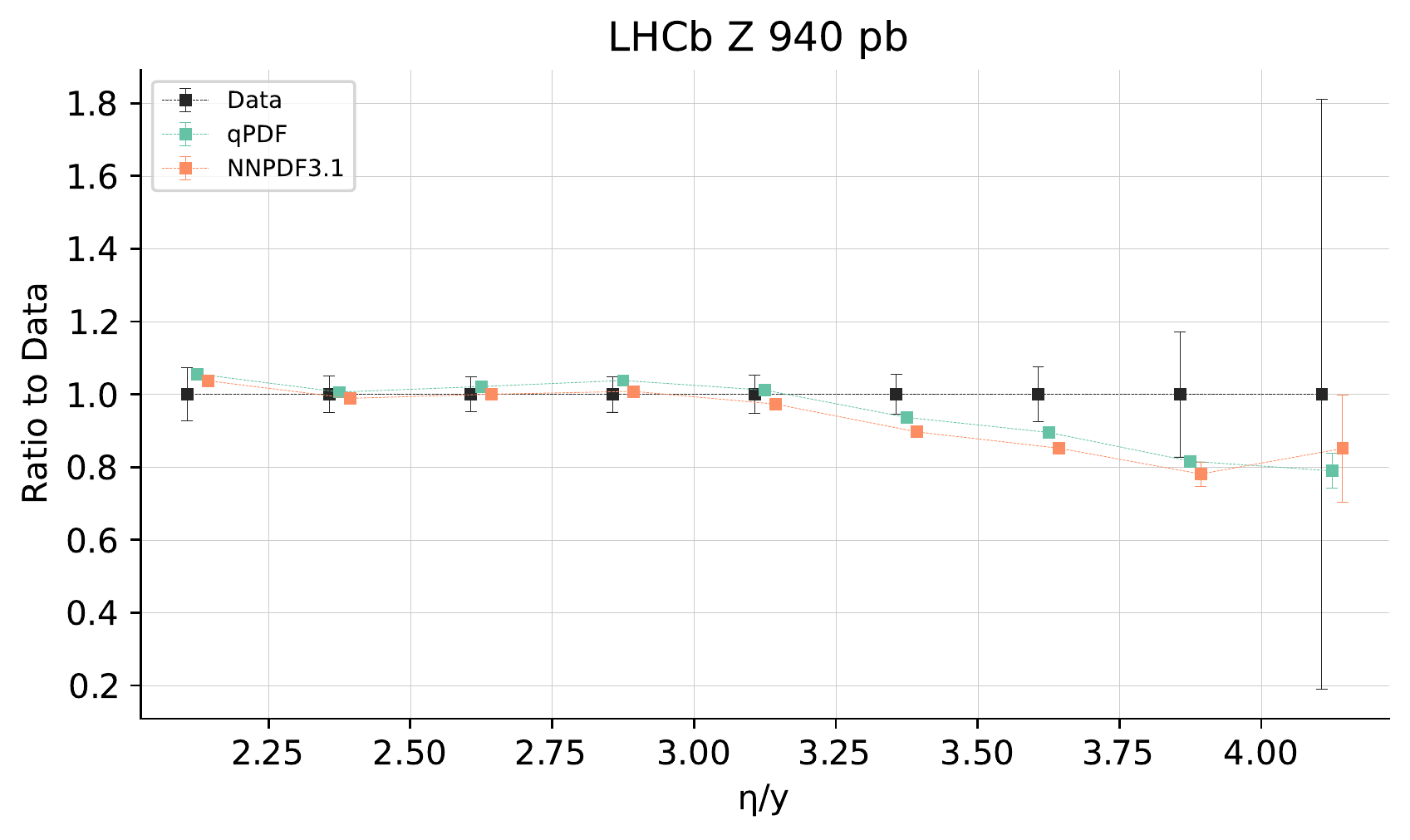}
    }
    \caption{Theoretical predictions computed with the method describe in~\cite{Bertone:2016lga} in order to compare
    the same prediction with three different PDF sets. We note that the predictions for the qPDF set is compatible with
both the experimental measurements and the released PDF set. The parton-level calculation has been performed with the
NLOjet++~\cite{Nagy:2001fj} and MCFM~\cite{Campbell:2019dru} tools.}\label{fig:expresults}
\end{figure*}

In Fig.~\ref{fig:expresults} we show specifically a comparison between the reference NNPDF3.1 and qPDF for selected
datasets, we also provide the LHAPDF-compatible PDF grid.
We observe that the accuracy of the qPDF central value is similar to that of NNPDF3.1.
Furthermore, the error bars for the predictions of both PDF set overlap with the experimental error bars, and, in some
cases, also among themselves.

\begin{figure}
  \includegraphics[width=0.5\textwidth]{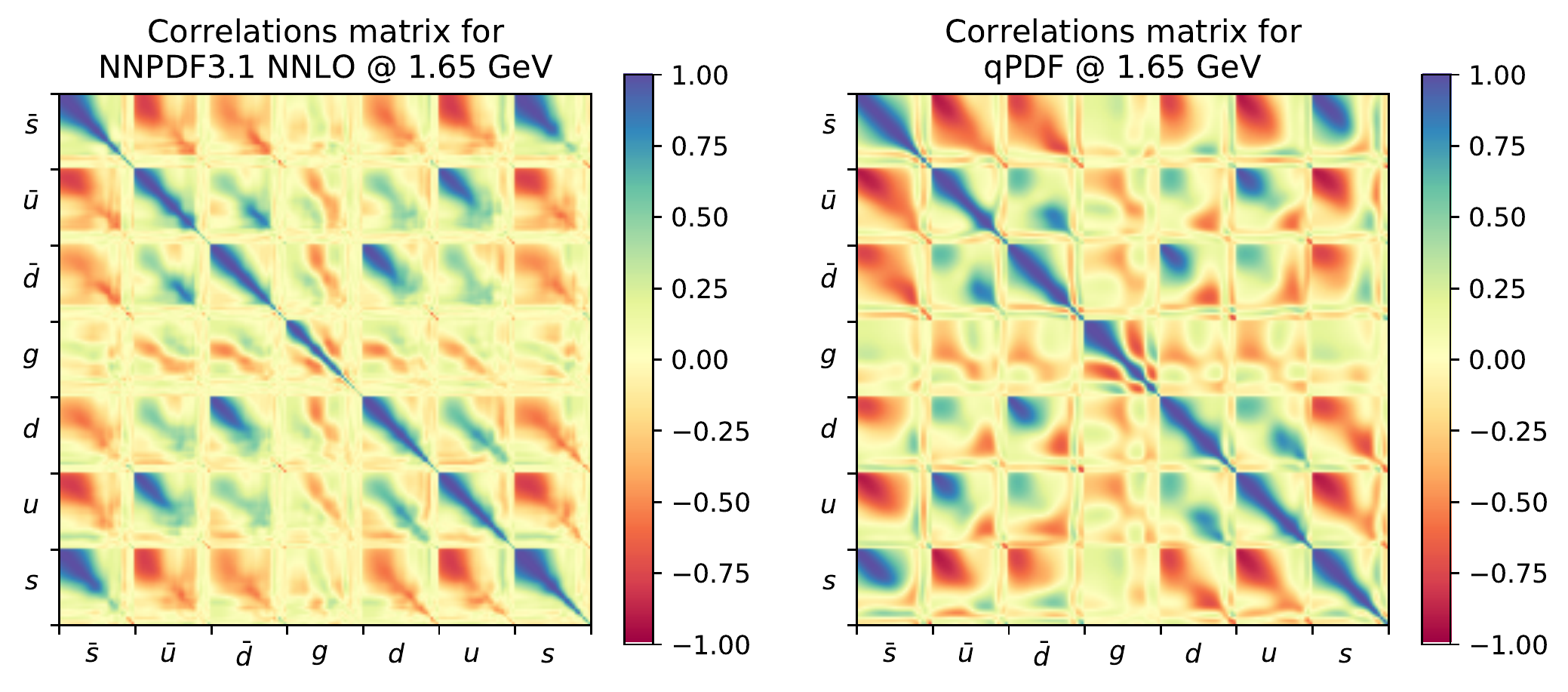}
  \caption{\label{fig:correlation}PDF correlation matrix for flavours in a grid
  of $x$ points for NNPDF3.1 NNLO (left) and the qPDF (right).}
\end{figure}

Finally, in Fig.~\ref{fig:correlation} we compute the PDF correlations for
NNPDF3.1 and qPDF replicas using Pearson's coefficient in a fixed grid of 100
points distributed logarithmically in $x=[10^{-4},1]$.

This leads us to conclude that the methodology described in this paper can be used for regression problems to unknown functional
forms such as the proton internal structure and produce results that are perfectly coherent, from a phenomenological
point of view, with the state of the art.
In addition we believe that with adequate tuning one could achieve the same level of accuracy of the classical
approach.

We finalize this section by showing phenomenological results where the LHAPDF grids produced with this approach are used
for a full fixed order prediction.
In summary going back circle to the master equation, i.e., computing numerically Eq.~\eqref{eq:convolution}
with no approximations using state of the art tools.

\section{Phenomenological results}\label{sec:pheno}

\begin{table}
  \begin{tabular}{c||cc}
  \hline
  Channel & NNPDF3.1 NNLO  & qPDF \tabularnewline
  \hline
  \hline
  $ggH$ & $31.04 \pm 0.30$ pb & $31.71 \pm 0.51$ pb \tabularnewline
  \hline
  $t\bar{t}H$ & $0.446 \pm 0.003$ pb & $0.464 \pm 0.008$ pb \tabularnewline
  \hline
  $WH$ & $0.133 \pm 0.002$ pb & $0.135 \pm 0.002$ pb \tabularnewline
  \hline
  $ZH$ & $0.0181 \pm 0.0002$ pb & $0.0184 \pm 0.0002$ pb \tabularnewline
  \hline
  VBF & $2.55 \pm 0.03$ pb & $2.62 \pm 0.04$ pb \tabularnewline
  \hline
  \end{tabular}
  \caption{\label{tab:higgs}The cross-sections for Higgs production at 13 TeV in
  various channels at NLO using the settings described in the text. From top to
  bottom: gluon fusion, $t\bar{t}H$ production, $WH$ production, $ZH$ production
  and vector boson fusion. We have assumed a Standard Model Higgs boson with
  mass $m_H=125$ GeV.}
\end{table}

In order to access the phenomenological implications of the qPDF fit, obtained
in the previous section, we compute and compare predictions for the most common
Higgs production channels.

The theoretical predictions are stored and computed with the
\texttt{PineAPPL}~\cite{Carrazza:2020gss,christopher_schwan_2020_3992765}
interface to \texttt{MadGraph5\_aMC@NLO}~\cite{Alwall:2014hca}. Cross-sections
have been computed for the LHC Run II kinematics, with a center-of-mass energy
of $\sqrt{s}=13$ TeV. In particular, we have generated NLO Higgs productions
tables for total cross-sections for gluon-fusion, vector-boson fusion,
associated production with $W$ and $Z$ bosons and associated production with top
quark pairs. No Higgs decays are included, since we are only interested in the
production dynamics. We have assumed a Standard Model Higgs boson with mass
$m_H=125$ GeV, and lepton cuts $p_{T,\ell} > 10$ GeV and $|\eta_{\ell}| < 2.5$.

In Table~\ref{tab:higgs} we present cross-section predictions for NNPDF3.1 NNLO
and qPDF. We observe that results are compatible and close to each other.

\section{Conclusion}\label{sec:conclusion}

In this work we proposed variational quantum circuit models for the
representation of PDFs in the context of high energy physics (HEP). We have
investigated and identified the most suitable Ansatz for the parametrization of
PDFs and defined a qPDF architecture. Using quantum circuit simulation on
classical hardware, we show that qPDFs are suitable for a global PDF
determination.

We highlight some advantages of the qPDF model when compared to the standard
machine learning methodology. Firstly, the availability of entanglement helps to
reduce the number of parameters required to obtain a flexible PDF
parametrization, in particular when compared to the number of parameters used by
an equivalent neural networks approach. Secondly, from a hardware implementation
point of view, the possibility to write the specific qPDF circuit in a quantum
processor, using its primitives (gates), will accelerate the evaluations and
training performance of PDFs. We expect that real quantum devices will be more
efficient in terms of energy power than classical hardware based on hardware
accelerators such as graphical process units (GPUs).

Furthermore, we propose a reconstruction method for evaluating the qPDF model in
a real quantum device using measurements. This procedure brings all the
difficulties that are typical of experimental quantum hardware, including noise,
error corrections and decoherence. The implementation of accurate and stable
qPDFs in a real quantum device still requires the development of hardware
architecture with lower gate error tolerances in comparison to the current
available machines.

On the other hand, our results should be considered as a proof-of-concept
exercise, given that the quantum simulation performance are still not
competitive with an equivalent machine learning implementation. The qPDF
approach may show advantages when more precise quantum devices will be
available.

Nevertheless, this is a first attempt to bridge the power of quantum machine
learning algorithms into the complexity of PDF determination. We auspicate that
the approach presented here will inspire new HEP applications which may benefit
from quantum computing.

\acknowledgments

We thank Jos\'e Ignacio Latorre, Sergi Ramos-Calderer, Carlos Bravo-Prieto and
Diego Garc\'ia-Mart\'in for useful and insightful discussions.
APS is supported by Projects PGC2018-095862-B-C22 and ERDF Operational Program of Catalunya QuantumCAT ref. 001-P-001644. APS acknowledges CaixaBank for its support of this work through Barcelona Supercomputing Center’s project CaixaBank Computación Cuántica.
JCM and SC are supported by the European Research Council under the European
Union's Horizon 2020 research and innovation Programme (grant agreement number
740006).
This work is supported by the Quantum Research Centre at the Technology
Innovation Institute.

\bibliography{blbl.bib}

\end{document}